\documentclass[12pt]{article}
\usepackage{amssymb}
\usepackage{amsmath}
\usepackage{amstext}
\usepackage{graphicx,epsfig}
\usepackage{epsfig}
\usepackage{verbatim} 
\usepackage{fancyhdr}
\usepackage{fancybox}
\usepackage{color}
\usepackage{ulem,bbold}
\usepackage{enumitem}
\usepackage{subfigure}
\usepackage{bbm}
\usepackage{parskip}

\newcommand{\Comment}[1]{{}}
\definecolor{MyDarkBlue}{rgb}{0.15,0.15,0.45}

\newcommand{\be}{\begin{equation}}  
\newcommand{\ee}{\end{equation}}  
\newcommand{\bea}{\begin{eqnarray}}  
\newcommand{\eea}{\end{eqnarray}}
\newcommand{\nn}{\nonumber}

\def\({\left(}
\def\){\right)}

\def\mpl{M_{\rm Pl}}
\def\p{\partial}
\def\mn{_{\mu \nu}}

\def\lsim{\mathrel{\rlap{\lower3pt\hbox{\hskip0pt$\sim$}}
     \raise1pt\hbox{$<$}}}         
\def\gsim{\mathrel{\rlap{\lower4pt\hbox{\hskip1pt$\sim$}}
     \raise1pt\hbox{$>$}}}         
\def\lsim{\mathrel{\rlap{\lower3pt\hbox{\hskip0pt$\sim$}}
     \raise1pt\hbox{$<$}}}         
\def\gsim{\mathrel{\rlap{\lower4pt\hbox{\hskip1pt$\sim$}}
     \raise1pt\hbox{$>$}}}         
\def\beq{\begin{eqnarray}}
\def\eeq{\end{eqnarray}}
\def\ba{\begin{eqnarray}}
\def\ea{\end{eqnarray}}
\def\({\left(}
\def\){\right)}

\numberwithin{equation}{section}

\begin{document}



\begin{center}
{\LARGE \bf{Inflation from Minkowski Space}}
\end{center} 
 \vspace{1truecm}
\thispagestyle{empty} 
\centerline{
David Pirtskhalava${}^{a,}$\footnote{E-mail address: david.pirtskhalava@sns.it}, 
Luca Santoni${}^{a,}$\footnote{E-mail address: luca.santoni@sns.it},
Enrico Trincherini ${}^{a,b,}$\footnote{E-mail address: enrico.trincherini@sns.it},
Patipan Uttayarat ${}^{c,d,}$\footnote{E-mail address: uttayapn@ucmail.uc.edu}
}

\vspace{0.5 cm}

\centerline{\it $^a$Scuola Normale Superiore, Piazza dei Cavalieri 7, 56126, Pisa, Italy}

 \vspace{.2cm}

\centerline{\it $^b$INFN - Sezione di Pisa, 56100 Pisa, Italy}

 \vspace{.2cm}

\centerline{{\it ${}^c$ 
Department of Physics, University of Cincinnati, Cincinnati, OH 45220 USA}}

 \vspace{.2cm}

\centerline{\it $^d$Department of Physics, Srinakharinwirot University, Wattana, Bangkok 10110 Thailand}

\begin{abstract}

We propose a class of scalar models that, once coupled to gravity, lead to cosmologies that smoothly and stably connect an inflationary quasi-de Sitter universe to a low, or even zero-curvature, maximally symmetric spacetime in the asymptotic past, strongly violating the null energy condition ($\dot H\gg H^2$) at intermediate times. The models are deformations of the conformal galileon lagrangian and are therefore based on symmetries, both exact and approximate, that ensure the quantum robustness of the whole picture. The resulting cosmological backgrounds can be viewed as regularized extensions of the \textit{galilean genesis} scenario, or, equivalently, as `early-time-complete' realizations of inflation.
The late-time inflationary dynamics possesses phenomenologically interesting properties: it can produce a large tensor-to-scalar ratio within the regime of validity of the effective field theory and can lead to sizeable equilateral nongaussianities.

\end{abstract}

\newpage

\thispagestyle{empty}
\newpage
\setcounter{page}{1}
\setcounter{footnote}{0}

\section{Introduction}
\parskip=5pt
\normalsize

The null energy condition (NEC) lies at the origin of the standard picture of early universe's cosmological evolution, determining many of its fundamental properties. For a universe dominated by a perfect fluid, satisfying the NEC is equivalent to the positivity of the sum of energy and pressure $\rho+p>0$, leading to ever-increasing energy density as the evolution is run backward in time. The regime of an $\mathcal{O}(1)$ sensitivity to the short-distance completion of gravitational interactions in the past is thus unavoidable for any NEC-satisfying cosmology.

Usually, violating the NEC is synonymous with instabilities -- at least for a system consisting of an arbitrary number of scalar fields with up to one derivative per field in the action \cite{Dubovsky:2005xd,Hsu:2004vr}. The theorem is not without loopholes, though. One possibility of evading it is provided by the \textit{ghost condensate} \cite{ArkaniHamed:2003uy}, that crucially relies on (spontaneously) broken Lorentz invariance in a way that gives rise to a non-standard $\omega \sim k^2$ infrared dispersion relation for the scalar driving the NEC violation. And indeed, it was argued in Ref. \cite{Creminelli:2006xe} that ghost condensation can lead to consistent alternative cosmologies with a weak ($\dot H\ll H^2$) violation of the null energy condition. Another loophole has emerged with the discovery of higher-derivative, yet ghost free scalar theory - the \textit{galileon} \cite{Nicolis:2008in}. The simplest such theory with a cubic self-interaction arises \cite{Luty:2003vm} in the context of the DGP model \cite{Dvali:2000hr}, while the full set of galileons have been found to describe the helicity-0 polarization of the graviton in dRGT theories of ghost-free massive gravity \cite{deRham:2010ik, deRham:2010kj}.

It has immediately been realized that (conformal) galileons can be implemented in building a NEC-violating alternative scenario to inflation, referred to as \textit{galilean genesis} (GG) \cite{Creminelli:2010ba}. In this class of models, conformal transformations (or, sometimes, just the dilatations \cite{Creminelli:2012my}) are assumed to be a symmetry of the flat-space theory, nonlinearly realized on a scalar field $\pi$, while couplings of $\pi$ to gravity are assumed to weakly break that symmetry. A crucial difference from inflation is that gravity is largely irrelevant for the early universe, described by GG: the cosmological phase of interest (during which the perturbations relevant for the CMB are produced) effectively takes place on a quasi-Minkowski spacetime, while scale-invariant density perturbations are naturally produced due to the unbroken dilatation invariance of the (time-dependent) scalar background\footnote{Similar ideas lie behind other constructions, such as that of a complex scalar rolling down a negative quartic potential \cite{Rubakov:2009np,Osipov:2010ee}, the \textit{pseudo-conformal universe} \cite{Hinterbichler:2011qk} and DBI genesis \cite{Hinterbichler:2012fr,Hinterbichler:2012yn}. The corresponding NEC-violating backgrounds are characterized by the same symmetry-breaking pattern, albeit technically realized in different ways.}. Moreover, flatness, homogeneity and horizon problems are automatically solved due to the quasi-Minkowski nature of the background spacetime and the gradual shrinking of the comoving Hubble horizon $(a H)^{-1}$. It is thus fair to say that, as far as the standard problems of the Big-Bang cosmology as well as density perturbations are concerned, galilean genesis is degenerate in its predictions with inflation. 

The differences come with the inclusion of tensor modes: irrelevance of gravity in genesis cosmologies results in a strongly blue-tilted and a completely unobservable (at least as far as the CMB experiments are concerned) spectrum of tensor perturbations \cite{Creminelli:2010ba}. For that reason, it is commonly believed that any possible detection of primordial gravitational waves (such as the one recently claimed by the BICEP2 collaboration \cite{Ade:2014xna}) would strongly disfavor genesis models, as well as their many variations. 
Indeed, a detectable, scale-invariant tensor spectrum requires that the background spacetime be (quasi-) de Sitter (dS) at the time of freezeout of the relevant set of modes (see, \textit{e.g.} \cite{Creminelli:2014wna} for a recent discussion). In the case that the interpretation of detected $B$-modes as a primordial signal persists, this would mean that any scenario that aims at describing the early universe should allow for a sufficiently extended period of de Sitter evolution. This apparently singles out the standard slow-roll inflation as the preferred paradigm for providing the flat and homogeneous universe with the particle horizon way beyond the observable patch. 

\begin{figure}
\includegraphics[width=0.7\textwidth]{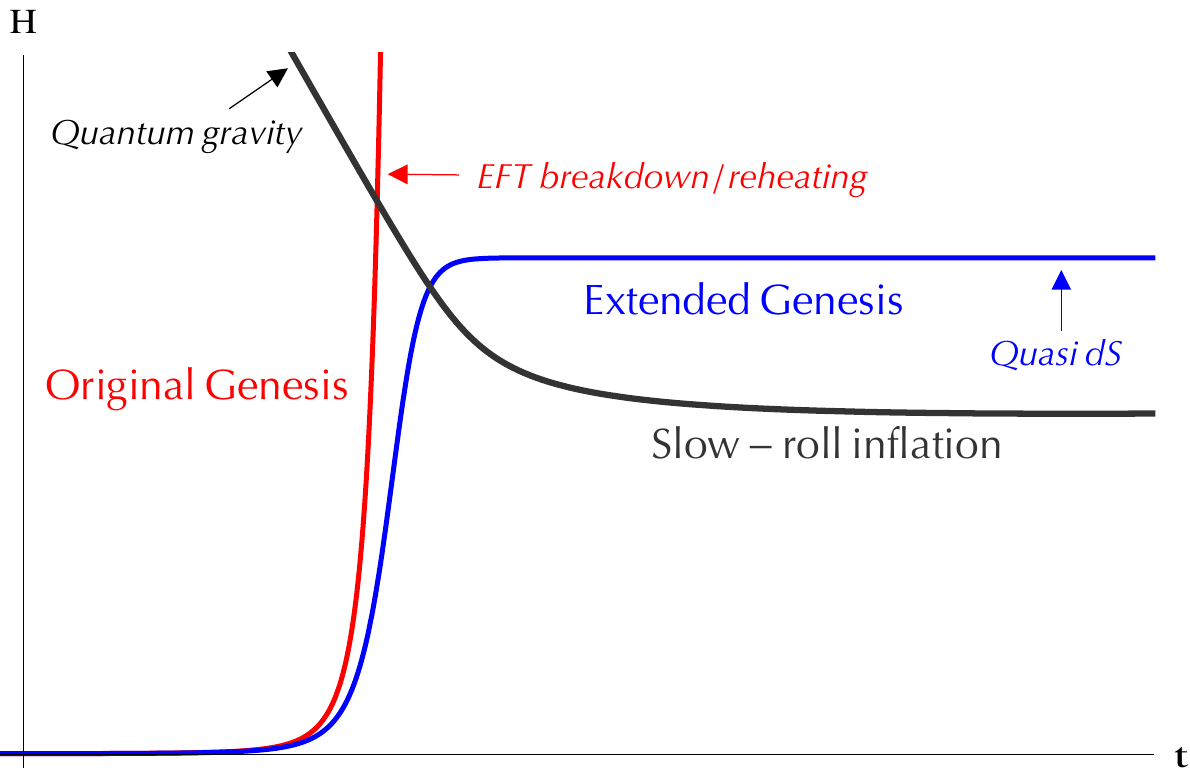}\centering
\caption{A sketch of the early universe's expansion rate as a function of time for the standard slow-roll inflation (black), as well as original (red) and extended (blue) genesis scenarios.}
\label{fig:1}
\end{figure}

One motivation of the present work is to re-assess the latter observation, with a focus on galilean genesis as an alternative to inflation. We will broadly define genesis as a phase of the universe with a strongly NEC-violating ($\varepsilon\equiv \dot H/H^2 \geq 1$) expansion that starts out in a low-curvature, maximally symmetric (essentially Minkowski or de Sitter) spacetime. Can such initial conditions result in a scale-invariant and unsuppressed tensor spectrum in a sufficiently broad range of physical scales? As noted above (at least for scalar-tensor theories we will be discussing below) generating scale-invariant tensor modes requires the geometry to be close to de Sitter for a certain period of time during the system's evolution. The question therefore reduces to that of the possibility for the universe to consistently evolve from a low/zero-curvature background in the far past to a much higher curvature inflationary dS spacetime capable of generating observable tensor spectrum at intermediate stages of its history. Because the system has to pass through a quasi de Sitter regime, one should be able to keep good theoretical control over the dynamics beyond the point when gravity starts playing a non-negligible role. Indeed, in the original GG, the moment of time $t_0$ at which gravity becomes order-one important is roughly the moment of the effective field theory (EFT) breakdown and not too long after that the universe is assumed to reheat, while all relevant cosmological perturbations are generated at times $t\ll t_0$ (we will assume time to flow from $t=-\infty$ towards $t=0$ throughout). This situation is sketched by the red curve on Fig. \ref{fig:1}. In terms of the model parameters, 
\beq
\label{tzero}
t_0\sim -\frac{f}{\mpl} \frac{1}{H_0}\, 
\eeq
where $f$ is the decay constant of $\pi$, while $H_0\ll f$ is a free parameter, setting the scale for the expansion rate around $t \sim t_0$ (the natural value for the decay constant is $f\sim \mpl$, which we will assume for definiteness in this section). The `slow-roll' parameter $\varepsilon$, starting out formally infinite at $t=-\infty$, decreases with time and is naively estimated to be of order unity at $t_0$. This means that the geometry can not be approximated by de Sitter space at any time during the genesis phase. 

While most of the qualitative features of GG directly follow from scale invariance of the (flat-space) $\pi$-lagrangian, the latter symmetry is badly broken by gravity around $t=t_0$. The background field value can be estimated at that time as
\beq
\phi\equiv e^\pi \simeq \mathcal{O}(1)~,
\eeq 
whereas throughout the genesis phase $\phi\ll 1$. One is then led to conclude that the loop-generated symmetry-breaking terms in the effective action for $\pi$ itself can start influencing the dynamics for $t\sim t_0$ -- even in the extreme case that these are down by the Planck scale. Indeed, the canonically normalized field $\pi_c$ becomes of order $\pi_c(t_0) \sim f $, making \textit{e.g.} the Planck-suppressed operator  $\pi_c (\p\pi_c)^2$ of the same order as the kinetic term. 
These estimates motivate extending the $\pi$ action by dilatation-breaking operators that, while irrelevant throughout the genesis phase, could in principle strongly influence the dynamics around the time when gravity becomes order-one important. 

We will show below that at least for a well-defined subclass of the resulting extensions, cosmological solutions do exist that, while resembling galilean genesis at early times, smoothly  extend \textit{beyond} the time $t=t_0$ as illustrated by the blue curve on Fig. \ref{fig:1}. These solutions asymptote, starting from some time $t_{i}$, to an inflationary (quasi) de Sitter space on which both the scalar and the tensor modes are generated with scale-invariant spectrum, just like in inflation. Nevertheless, the scenario at hand -- referred to as \textit{extended genesis} (EG) below -- crucially differs from inflation in that the universe's evolution at early times ($t\ll t_i$) looks nothing like that of the standard NEC-satisfying slow-roll models. Most importantly, NEC violation provides a possibility to avoid the singularity in the past, with the universe gradually relaxing to a low- (or even zero-) curvature space as it is run backwards in time. Due to the latter property, extended genesis can be alternatively viewed as a `UV' (or, to be more precise, as an early-time) -complete realization of inflation.

In the cases we consider below, the late-time dynamics of EG will be described by NEC-violating versions of \textit{galileon inflation} (also referred to as \textit{G-inflation}) \cite{Kobayashi:2010cm} -- a model that possesses a number of phenomenologically attractive properties. First, it can produce a large tensor-to-scalar ratio without trans-Planckian field excursions, unlike the standard slow-roll inflation \cite{Lyth:1996im} (because of the shift symmetry, the inflaton itself is not an observable in galileon inflation). Second, similar to ghost \cite{ArkaniHamed:2003uz, Senatore:2004rj} and DBI \cite{Alishahiha:2004eh} models, galileon inflation can lead to a sizeable equilateral nongaussianity. Finally, since $\pi$ itself acquires a scale-invariant spectrum, it is in principle unnecessary to invoke spectator fields (required in many alternatives to inflation) for generating the observed density perturbations. 

The paper is organized as follows. We start in Sec.\ref{sec2} by spelling out general criteria that a theory, capable of describing the genesis -- de Sitter transition of Fig. \ref{fig:1}, should satisfy. In the same section we give a simple example of a solution with the given feature. Sections \ref{ggal} and \ref{analytic} deal with an analytic construction of such theories, providing explicit examples of completely stable cosmological solutions exhibiting extended genesis. In Sec.\ref{hdim} we study possible effects of higher derivative operators in the effective theory on the scalar spectrum of the backgrounds under consideration. Finally, in Sec.\ref{concl} we conclude. Technical details, that would overwhelm the main body of the text, are collected in the two appendices.

The theories described in the rest of the paper are only intended as a starting point for constructing realistic early universe cosmologies based on EG. While we do touch on this in what follows, a fully realistic model-building is left for future work. Most importantly, however, our examples serve as a proof of principle of the possibility to smoothly and stably connect the inflationary quasi-de Sitter universe to a low or even zero-curvature, maximally symmetric spacetime in the asymptotic past.

\section{Generalities}
\label{sec2}

Before diving into a more detailed discussion, we briefly highlight the major properties of theories allowing for the genesis - dS transition. We expect these properties to be the defining ingredient of any other construction capable of achieving our goals. Most importantly, the theories of interest enjoy an enhanced symmetry both for small as well as for large values of the 'sigma model' field $\phi$. In both limits $e^\pi\ll 1$ and $e^\pi\gg 1$, the (flat-space) $\pi$-lagrangian will acquire invariance either under dilatations
\beq
\label{dil}
\pi(x)\to \pi(e^\lambda x)+\lambda~,
\eeq
describing the scale-invariant (and, in special cases, conformal) galileon \cite{Nicolis:2008in}, 
or under constant \textit{shifts}
\beq
\label{shift}
\pi(x)\to \pi(x)+\lambda~,
\eeq
describing $P(X)$ or ordinary galileon-type theories\footnote{By `$P(X)$ theories' we mean theories, defined by their lagrangian being an arbitrary function $P$ of the combination $X\equiv -(\p\pi)^2$. In the inflationary context these were first studied in \cite{ArmendarizPicon:1999rj}. } with ghost condensation, see \textit{e.g.} \cite{ArkaniHamed:2003uy,ArkaniHamed:2003uz}. Apart from the two (asymptotically) exact symmetries, for $e^\pi\gsim 1$ the theories under consideration will be \textit{approximately} invariant under internal galilean transformations, 
\beq
\label{galinv}
\pi\to\pi+b_\mu x^\mu~,
\eeq
with $b_\mu$ a constant four-vector. The reason it is useful to think of galilean invariance as an approximate symmetry is that the operator that breaks it has a parametrically suppressed Wilson coefficient in the effective theory. Approximate invariance under \eqref{galinv} then makes this suppression stable under loop corrections, see the discussion below. Galilean invariance becomes more and more pronounced as $\pi\to 0$.
As we will see in Sec.\ref{analytic}, in certain cases the small-field regime will itself consist of two qualitatively different stages -- the system gradually evolving from ghost condensate (described by an effectively shift-symmetric theory)  in the asymptotic past, into galilean genesis with an enhanced scale invariance \eqref{dil} -- all while $\phi\ll 1$.

The asymptotically emergent symmetries are precisely what makes the existence of NEC-violating cosmologies, interpolating between Minkowski and de Sitter spacetimes possible. Let us \textit{e.g.} consider the genesis-de Sitter transition of Fig. \ref{fig:1}. The enhanced conformal invariance at early times/small field values\footnote{To avoid confusion, we note again that `small field values' refers to the expectation value of the sigma model field $e^\pi$, while the goldstone $\pi$ is characterized by large negative values in the given regime. } \textit{generically} gives rise to  galilean genesis-like evolution of the universe, whereby conformal invariance, $SO(4,2)$, gets broken down to the maximal de Sitter subgroup $SO(4,1)$ by a time-dependent $\pi$-background\footnote{We stress that while de Sitter group is the (linearly realized) symmetry group of the scalar action, the geometry throughout the galilean genesis phase remains close to flat.} -- the Hubble rate and the sigma model field $e^\pi$ growing as time flows from $t=-\infty$ towards $t=0$. Whenever $e^\pi$ starts exceeding unity on the other hand, the emergent shift symmetry naturally leads to an attractor solution with de Sitter \textit{geometry} on which the scalar acquires a linear profile, $\pi\propto t$ \cite{ArkaniHamed:2003uz, Kobayashi:2010cm}. This qualitatively explains the gradual transformation between genesis and de Sitter phases as illustrated by the blue curve in Fig. \ref{fig:1}. 

Last but not least, the enhanced symmetries for large and small field values lead to the quantum robustness of the whole qualitative picture. Indeed, both symmetries \eqref{dil} and \eqref{shift} are broken at order one when $e^\pi\sim 1$, making it hard to argue in favor of quantum stability of the detailed intermediate-time behaviour of our solutions. Nevertheless, the scale and shift symmetries are fully intact asymptotically, determining radiative stability of both the early- and the late-time dynamics. Backgrounds exhibiting the genesis-de Sitter transition can thus be expected to exist \textit{generically}, since both of the asymptotic solutions arise solely from symmetry considerations. A similar discussion of quantum robustness has been given in Ref. \cite{Elder:2013gya} in the context of flat-space constructions interpolating between NEC-satisfying and NEC-violating vacua.

For completeness, in the rest of the section we give a relatively detailed overview of the two asymptotic regimes of the solutions we wish to study.

\subsection*{Galilean genesis}

Conformal symmetry, $SO(4,2)$, can be \textit{generically} broken down to its maximal, de Sitter subgroup $SO(4,1)$ by a time-dependent scalar profile \cite{Fubini:1976jm, Nicolis:2008in,Nicolis:2009qm}. One way to achieve such  breaking is via the (simplest non-trivial) \textit{conformal galileon} lagrangian
\beq
\label{gg}
S_{\text{1}}=\int d^4 x \sqrt{-g}~\bigg[ f^2 e^{2\pi}(\p\pi)^2+\frac{f^3}{\Lambda^3}(\p\pi)^2\Box\pi+\frac{f^3}{2\Lambda^3} (\p\pi)^4\bigg ]~.
\eeq
It can be straightforwardly checked that the theory possesses an exact rolling solution on flat spacetime \cite{Nicolis:2009qm,Creminelli:2010ba}
\beq
\label{ggsol}
e^\pi=-\frac{1}{H_0 t}, \qquad H_0^2=\frac{2\Lambda^3}{3f}~,
\eeq
leading precisely to the $SO(4,2) \rightarrow SO(4,1)$ breaking pattern. The dilatation invariance, left unbroken by the background, leads to vanishing of its energy density\footnote{This immediately follows from scale invariance ($\rho\propto \frac{1}{t^4}$) plus the energy conservation ($\dot\rho=0$).}, $\rho=0$, while the pressure $p=-2f^2/(H_0^2t^4)$ is negative -- implying a strongly NEC-violating ($\dot H\gg H^2$) expansion \cite{Creminelli:2010ba}. The universe described by GG starts out in flat spacetime, the Hubble rate growing according to the second Friedmann equation
$2\mpl^2 \dot H=-(\rho+p)$, which upon integration yields
\beq
\label{ggsol1}
H\simeq-\frac{1}{3}\frac{f^2}{\mpl^2}\frac{1}{H^2_0t^3}~.
\eeq
The time $t_0$ at which gravity starts playing non-negligible role ($H\sim \dot\pi$), can be estimated as in \eqref{tzero}. It roughly coincides with the time of EFT breakdown/start of reheating. Scalar perturbations, relevant for the CMB are instead produced at earlier times $t\lsim t_0$, via minimally coupling an additional, scaling dimension-0 field $\varphi$ to the 'fake de Sitter' metric $g^{\text{dS}}_{\mn}=e^{2\pi}\eta_{\mn}$. This leads to a scale-invariant spectrum for the spectator $\varphi$ (despite the background metric being practically flat), that can be later imprinted on the physical curvature perturbation $\zeta$ through one of the standard mechanisms \cite{Enqvist:2001zp,Lyth:2001nq,Dvali:2003em}. The near-to-flat geometry on the other hand implies a strongly blue-tilted tensor spectrum $P_h(k)\sim k^2$, largely irrelevant for CMB observations \cite{Creminelli:2010ba}. 

\subsection*{Galileon inflation}

An immediate candidate for describing the late-time de Sitter asymptotics of the solutions of interest is a cubic galileon theory with a small quartic self-interaction, defined by the following action
\beq
\label{ginf}
S_{\text{2}}=\int d^4 x \sqrt{-g} ~\bigg[f^2 (\p\pi)^2+\gamma_3\frac{f^3}{\Lambda^3}(\p\pi)^2\Box\pi+\gamma_4\frac{f^3}{2\Lambda^3} (\p\pi)^4\bigg ]~,
\eeq
where $\gamma_{3,4}$ are constant parameters. The form of the above action is dictated by the early-time genesis asymptotics. Indeed, both of the interactions in \eqref{ginf} are also present in \eqref{gg}, the only difference between the two theories being that the former lacks scale invariance. Moreover, the galileon term will be crucial for the speed of sound in the inflationary regime to be strictly positive. 

Inflationary solutions in this theory have been studied in Ref. \cite{Kobayashi:2010cm}. Here we will re-derive all of the (qualitative) results of the latter reference using simple EFT considerations. In addition, we will provide arguments in favor of the quantum robustness of these results -- something that, to the best of our knowledge, has not been pointed out before. 

The Friedmann equation and the equation of motion for $\pi$ take on the following form on spatially flat FRW backgrounds 
\begin{gather}
H^2=\frac{f^2}{3 \mpl^2 H_0^2} ~\( \gamma_4 \dot\pi^4+4\gamma_3 H\dot\pi^3-H_0^2\dot\pi^2    \)~, \\
\( 4 \gamma_4\dot\pi^2 +8\gamma_3 H\dot\pi -2 H_0^2\)\ddot\pi +4 \gamma_4 H\dot\pi^3+4\gamma_3\(3 H^2+\dot H\)\dot\pi^2-6 H_0^2 H\dot\pi=0 ~,
\end{gather}
making existence of de Sitter vacua ($H=\text{const}$) with a linear $\pi\propto t$ profile explicit -- a direct consequence of shift-invariance \eqref{shift} of the $\pi$-lagrangian. 
Furthermore, the expansion rate and the scalar profile can be estimated as
\beq
\label{sol1}
H^2\sim \frac{f^2}{\mpl^2} H_0^2~,\qquad \dot \pi\sim H_0~.
\eeq

The simplest and the most straightforward way of studying the spectrum of scalar perturbations is based on the effective theory of inflation \cite{Creminelli:2006xe,Cheung:2007st}. The formalism is reviewed in great detail in Appendix \ref{appA} and in Sec.\ref{hdim}, so we will content ourselves with a brief treatment here. The two operators in the effective theory that lead to non-trivial dynamics at high energies are the $\delta N^2$ and $\delta N\delta E^i_{~i}$ terms in the notation of Eq.\eqref{s_pi}. The coefficients of these terms, given for a generalized theory of Sec.\ref{ggal} in Eq.\eqref{M's}  (the present case corresponds to simply setting $\mathcal{F}_2=1$ in the latter expressions), are of order 
\beq
\label{sol2}
M^4\sim f^2 H_0^2, \qquad \hat M^3_3\sim f^2 H_0~.
\eeq
A particularly useful regime of the system is the one corresponding to the short-distance, \textit{decoupling} limit, that allows to zoom onto the relevant high-energy degrees of freedom present in the theory\footnote{Although a `short distance' limit, the decoupling limit is crucially valid at distances parametrically greater than the inflationary Hubble scale at which the scalar spectrum is evaluated.}. In this limit the dynamics of the scalar is fully captured by the Goldstone mode corresponding to the breaking of time translation-invariance, which we will refer to as $\pi_g$ (we will rely on the reader to not confuse the Goldstone boson with the fundamental galileon field $\pi$). The decoupling limit action for $\pi_g$, assuming $\mpl^2 \dot H\ll f^2 H_0^2$, reads \cite{Creminelli:2006xe,Cheung:2007st}
\beq
\label{goldst}
S=\int d^4 x \sqrt{-\bar g}\bigg [M_0^4\(\dot \pi_g^2-c_s^2 \frac{(\nabla\pi_g)^2}{a^2}\)-M_0^4 \dot\pi_g\frac{(\nabla\pi_g)^2}{a^2}+\frac{\hat M^3_3}{2}\frac{\nabla^2\pi_g (\nabla\pi_g)^2}{a^4}+\dots\bigg],~~~~~
\eeq
where $\bar g$ is the unperturbed de Sitter metric, and we have made use of the following notation
\beq
\label{cssq}
M^4_0=\frac{M^4}{2}-3\hat M^3_3 H\sim f^2 H_0^2, \qquad c_s^2=\frac{3 \hat M^3_3 H}{M_0^4}~.
\eeq
Recalling that the physical curvature perturbation is related to $\pi_g$ by a gauge transformation, $\zeta=-H\pi_g$, one can directly read off the expression for the power spectrum of scalar perturbations from the Goldstone action \eqref{goldst}
\beq
\langle\zeta_{\vec k_1} \zeta_{\vec k_2} \rangle=(2\pi)^3 \delta(\vec k_1+\vec k_2) \frac{1}{k_1^3}~\frac{H^4}{M_0^4 c_s^3}~,
\eeq
where all quantities on the right hand side are assumed to be evaluated at horizon crossing $k_1=a H$, as usual. The tensor spectrum on the other hand is given by the universal formula $\Delta^2_\gamma\sim H^2/\mpl^2$. Using Eqs.  \eqref{sol1} and \eqref{sol2}, as well as Eq.\eqref{cssq} for the speed of sound, one finds the following expressions for the dimensionless power spectra
\beq
\Delta^2_\zeta \sim \frac{f^{1/2} H_0^2}{\mpl^{5/2}},\qquad \Delta^2_\gamma \sim \frac{f^2 H_0^2}{\mpl^4}, \qquad r=\frac{\Delta^2_\gamma}{\Delta^2_\zeta } \sim \(\frac{f}{\mpl}\)^{3/2}~,
\eeq
in agreement with the results of \cite{Kobayashi:2010cm} (see the latter reference for the computation in the full theory, including the precise numerical factors). Moreover, one can see from the above that the tensor-to-scalar ratio can easily be made large enough to be detectable if $f$ is sufficiently close to $\mpl$ -- all within the regime of validity of the underlying effective field theory. 

One can go further and estimate the amount of non-Gaussianity that the model under consideration is expected to generate. The most relevant cubic interactions of $\pi_g$, giving the leading non-Gaussian effects have been explicitly written out in \eqref{goldst}. The three-point function is of the equilateral shape for both of these \cite{Bartolo:2010bj} (see also \cite{Kobayashi:2011pc}). The amplitude on the other hand can be estimated e.g. for the $\dot \pi_g (\nabla\pi_g)^2$ operator in the standard way \cite{Cheung:2007st}  (again, all terms on the r.h.s. should be understood as evaluated at horizon-crossing)
\beq
f_{NL}\sim \frac{1}{\zeta} \frac{\mathcal{L}_{\dot \pi_g (\nabla\pi_g)^2}}{\mathcal{L}_{\dot \pi_g^2}}\sim \frac{1}{H \pi}\frac{H\(k/a\)^2\pi}{H^2}=\frac{1}{c_s^2}~.
\eeq
This leads to the amount of non-Gaussianity similar to that in DBI models of inflation \cite{Alishahiha:2004eh}.
An analogous estimate shows that the second cubic self-interaction generates a comparable contribution to $f_{NL}$. 

\begin{figure}
\centering
\includegraphics[width=.4\textwidth]{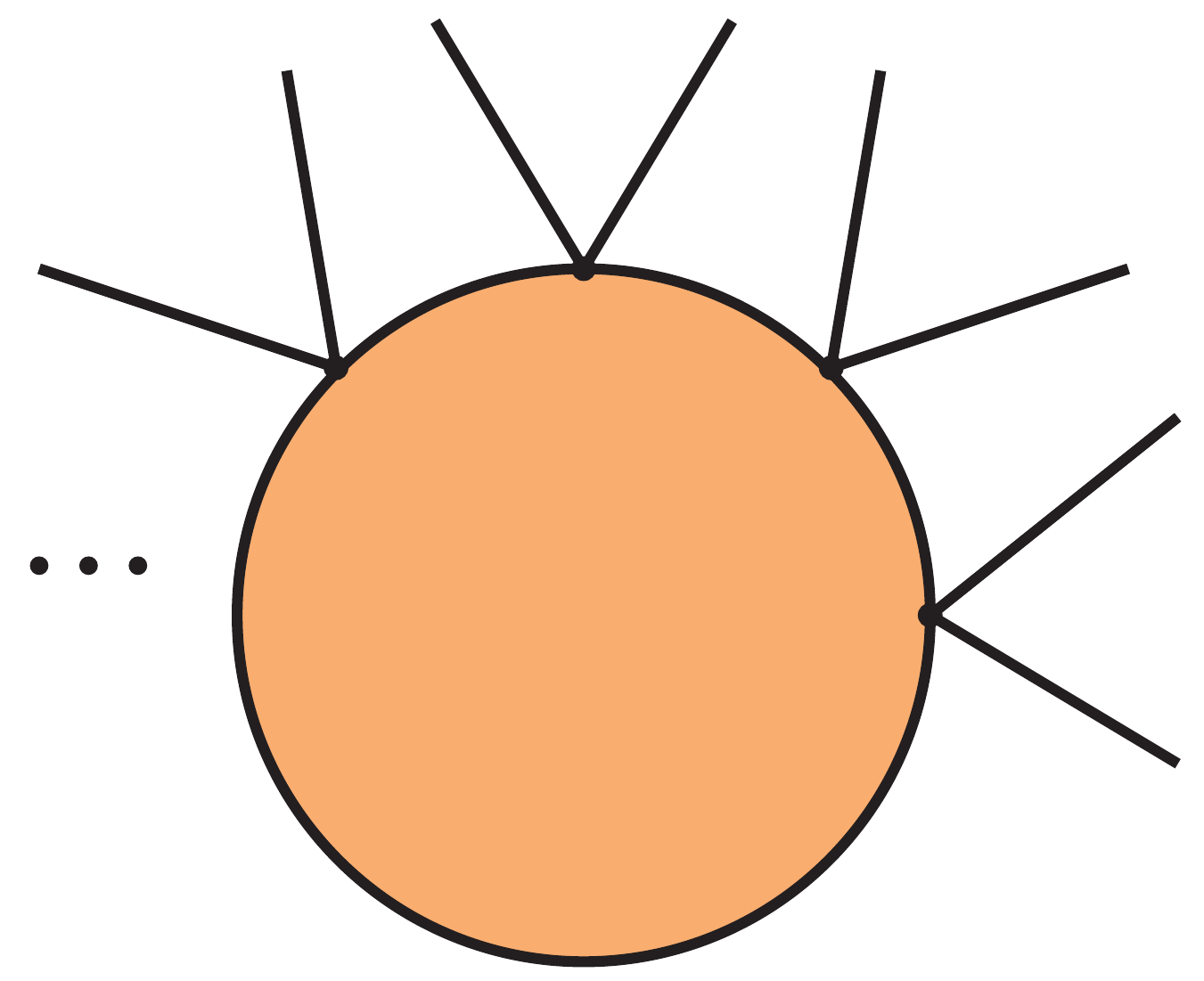} 
\caption{The diagram, responsible for the dominant quantum correction to the background solution in galileon inflation.}
\label{diagram}
\end{figure}

At this point one may be worried about the UV sensitivity of the obtained inflationary solution, since the scales suppressing the two (canonically-normalized) interactions in \eqref{ginf}, $\Lambda$ and $\tilde \Lambda\equiv (f\Lambda^3)^{1/4}$,  are parametrically separated\footnote{The same is true for the conformally invariant theory \eqref{gg}, however there the hierarchy is not a problem, as it is completely stabilized by conformal symmetry.} ($\tilde\Lambda\gg \Lambda$). In fact, this separation is crucial if all three terms in the lagrangian are to play an equally important role on the given background; indeed, for $f\sim\mpl$, one can estimate the magnitude of each operator (including the kinetic term) to be of order $\rho_{dS}\equiv f^2 H_0^2$. Interpreting $\Lambda$ -- the smallest of the two scales -- as the quantum cutoff of the theory then, nothing apparently prevents a loop-generated self-interaction \textit{e.g.} of the form  $(\p\pi)^4/\Lambda^4$, which would parametrically dominate over the last term (and therefore over all terms) in \eqref{ginf}. This would impair the whole description of the obtained dS backgrounds. Fortunately, the latter reasoning turns out to be too hasty and the background can in fact be trusted. This can be seen as follows. Consider all loop diagrams, generating a term of the form $(\p\pi)^{2n}$. What is the smallest scale that can suppress such an operator? To answer this question, we note that whatever the diagram responsible for this operator is, it can not have an external leg originating from the cubic galileon vertex, since this would lead to at least two derivatives acting on the corresponding asymptotic state (the reason for this lies in the non-renormalization theorem that severely constrains the form of quantum corrections in galileon theories \cite{Luty:2003vm}). We thus conclude that all external legs in the diagram originate from the quartic interaction, which introduces a suppression of \textit{at least one} factor of $f$ per pair of fields in the corresponding effective vertex. The least suppressed loop corrections of the given form thus correspond to the diagram of  Fig. \ref{diagram}. Assuming that all the rest of the vertices are those of the cubic galileon (and therefore only introduce factors of $\Lambda$, but not of $f$) and that loop integrals are cut-off at energies of order the strong-coupling scale of the theory $\Lambda$, one arrives at the following conservative estimate for the magnitude of the operators of the given type
\beq
\mathcal{L}_{loop}= \frac{(\p\pi)^{2n}}{f^n \Lambda^{3k}}, \qquad k=n-\frac{4}{3}~.
\eeq
Evaluating $\mathcal{L}_{loop}$ on the classical de Sitter background gives
\beq
\mathcal{L}_{loop}=f^{4/3} H_0^{8/3}\ll \rho_{ds}~,
\eeq
independently of $n$. This leads one to conclude that quantum corrections of the form $(\p\pi)^{2n}$ do not modify the background obtained from the lagrangian (\ref{ginf}). Note that the non-renormalization properties of the galileon play a crucial role in the latter conclusion. Furthermore, the fact that $\pi$ acquires a linear profile on de Sitter backgrounds makes operators with more than one derivative per field similarly irrelevant, since they are suppressed by powers of the scale $H_0$, parametrically smaller than $f$ and $\Lambda$.

The $\pi\propto t$ solution describes a perfect de Sitter space, leading to exactly scale-invariant perturbations; adding a small potential (or deforming the form of the action otherwise), both the scalar and the tensor modes can be produced with slightly tilted spectra -- just as they are in the canonical inflationary case. In addition, to complete the picture one of course has to specify a mechanism for exiting the de Sitter phase. There are known ways of achieving this, and we refer the interested reader to works, dealing with similar issues in various contexts \cite{ArkaniHamed:2003uz, Senatore:2004rj, Osipov:2010ee, Ivanov:2014yla}.

\subsection*{An explicit example}
As a simple example of a theory with the above-described asymptotics, one can consider the deformed galilean genesis lagrangian
\ba
S=\int d^4x~ \sqrt{-g} ~\bigg[\frac{1}{2}\mpl^2 R +f^2~ \frac{e^{2\pi}}{1+\beta e^{2\pi}} ~(\p\pi)^2+\frac{f^3}{\Lambda^3}(\p\pi)^2\Box\pi +\frac{f^3}{2\Lambda^3}(\p\pi)^4  \bigg]~,
\label{simplemodel}
\ea 
with $\beta$ an arbitrary constant. For $\beta=0$ the theory is just the conformal galileon and when starting out in the GG phase, the expansion rate of the universe  diverges and the background exits the regime of validity of the EFT at some finite time (see the red curve in Fig. \ref{fig:1}) -- the scalar profile growing as $e^\pi\sim 1/t$ throughout. For a nonzero $\beta$ on the other hand, the dynamics of the system is completely altered as soon as $\beta e^{2\pi}$ becomes of order, or greater than one: the theory becomes effectively described by a $P(X)$ - type lagrangian with a cubic galileon self-interaction, resulting in transition into an inflationary de Sitter phase. 
The corresponding solutions are studied in Appendix \ref{num}, where the existence of extended genesis cosmologies is illustrated via numerical analysis: the system clearly exhibits transition from genesis into a quasi - de Sitter regime precisely around the time $t_0$ given in \eqref{tzero}, see Fig.  \ref{figuretwo}. Perhaps the only downside of this simple model is the short temporal region with gradient instability at intermediate times: while completely free from ghosts, the squared speed of sound of the scalar perturbation goes slightly negative on the given background around $t\sim t_0$ for a period of roughly a Hubble time, as shown in Fig. \ref{figuretwo} (we will track down the origin of the gradient instability analytically  in Sec. \ref{analytic} ). While certainly a problem in the classical theory, higher-order effects can in principle take care of this issue -- rendering the cosmological evolution free from instabilities, see the discussion in Sec. \ref{hdim} and Appendix \ref{num}.

We refer the reader to Appendix \ref{num} for a detailed discussion of numerical solutions to the illustrative model \eqref{simplemodel}, and turn to a systematic construction of theories leading to early universe cosmology with the genesis - dS transition in the next section.

\section{Generalized galileons}
\label{ggal}

In the present and the next sections we will take on the task of obtaining (analytic) cosmological solutions exhibiting extended genesis. Rather than constructing solutions to a particular theory obeying the asymptotic scale and shift symmetries described in Sec. \ref{sec2}, we will employ the trick used in Ref. \cite{Elder:2013gya}, where the appropriate theory itself is inverse - engineered based on a postulated ansatz for the desired cosmological solution. The asymptotic symmetries, as we will see, then follow automatically from the construction which we describe in what follows.

Consider a (generally dilatation-breaking) deformation of the galilean genesis lagrangian
\beq
\label{ggg}
\mathcal{S}_\pi=\int d^4 x ~\sqrt{-g} ~\bigg[f^2 \mathcal{F}_1(\pi) (\p\pi)^2+\frac{f^3}{\Lambda^3} (\p\pi)^2\Box\pi +\frac{f^3}{2\Lambda^3} \mathcal{F}_2(\pi) (\p\pi)^4    \bigg ]
\eeq 
where $\mathcal{F}_{1,2}$ are \textit{a priori} arbitrary dimensionless functions of the galileon field $\pi$. We will interchangeably use the two scales $\Lambda$ and $H_0$ (as defined in \eqref{ggsol}) throughout. The dynamics of the system is governed by the Einstein's equations plus the scalar equation of motion. These however are not independent: as a consequence of diffeomorphism invariance, the scalar equation can be traded for the conservation of its stress-energy tensor via
\beq
\label{s.eom}
\nabla_\mu T^\mu_{~\nu}=-\frac{\delta S}{\delta \pi} \p_\nu\pi~.
\eeq
On homogeneous FRW backgrounds, it is the energy conservation, $\dot \rho + 3H (\rho+p)=0$, that yields the $\pi$ equation of motion. Energy conservation on the other hand follows from the temporal and space components of the Einstein's equations -- therefore we can choose the latter two to make up a complete system determining background evolution. 
The stress-energy tensor, sourced by $\pi$ in \eqref{ggg} is  
\bea
T^\pi_{\mn}=&-&f^2 \mathcal{F}_1(\pi) ~[2\p_\mu\pi\p_\nu\pi-g_{\mn} (\p\pi)^2]\nn \\
&-&\frac{f^3}{\Lambda^3}~[2\p_\mu\pi\p_\nu\pi\Box\pi-
\p_\mu\pi\p_\nu(\p\pi)^2-\p_\nu\pi\p_\mu(\p\pi)^2
+g_{\mn}\p_\lambda\pi\p^\lambda(\p\pi)^2]\nn \\
&-&\frac{f^3}{2 \Lambda^3}~\mathcal{F}_2(\pi) ~[4 (\p\pi)^2\p_\mu\pi\p_\nu\pi-g_{\mn} (\p\pi)^4]~,
\ea
leading to the following expressions for the energy density and pressure due to a homogeneous $\pi$-profile
\bea
\label{rho}
\rho &=&\frac{f^2}{H_0^2} ~\dot \pi^2 \big [ \mathcal{F}_2(\pi)\dot\pi^2+4 H\dot\pi -  H_0^2\mathcal{F}_1(\pi)\big ] ~,\\
\label{press}
p &=& \frac{f^2}{3 H_0^2}~\dot\pi^2 \bigg[ \mathcal{F}_2(\pi) \dot\pi^2-4\ddot \pi-3H_0^2\mathcal{F}_1(\pi) \bigg]~.
\ea
The two functions $\mathcal{F}_{1,2}(\pi)$ can be solved for with the help of the temporal and spatial components of Einstein's equations, $3 \mpl^2 H^2=\rho$ and $\mpl^2 (3 H^2+2 \dot H)=-p$, which yields
\beq
\label{f1}
\mathcal{F}_1&=&\frac{6 \mpl^2 H_0^2 H^2+3\mpl^2 H_0^2\dot H-2 f^2 H\dot \pi^3-2f^2 \dot\pi^2\ddot\pi}{f^2 H_0^2\dot\pi^2}\\
\label{f2}
\mathcal{F}_2&=&\frac{9\mpl^2 H_0^2 H^2+3\mpl^2 H_0^2\dot H-6 f^2 H\dot \pi^3-2f^2 \dot\pi^2\ddot\pi}{f^2 \dot\pi^4}~.
\eeq  
Now, for any \textit{postulated} homogeneous profile of the scalar and the Hubble rate, one can find the theory (i.e. find $\mathcal{F}_{1,2}(\pi)$) such that the desired background solves its equations of motion. The recipe for constructing the relevant solutions is given as follows:
\begin{itemize}
\item Postulate background profiles $\pi_0(t)$ and $H(t)$~

\item For the chosen background solutions, find the time-dependent functions $\mathcal{F}_{1,2} (t)$  with the help of \eqref{f1} and \eqref{f2} 

\item Invert the expression for $\pi_0(t)$ to find $t=t(\pi_0)$

\item Using the previous steps, find $\mathcal{F}_{1,2}$ as functions of $\pi_0$: $\mathcal{F}_{1,2}=\mathcal{F}_{1,2}\(t(\pi_0)\)$~.

\end{itemize}
That way one can formally construct theories admitting arbitrary cosmological profiles for $\pi$ and $H$. Although such an \textit{ad hoc} construction might look uncomfortable, we will see that at least for the solutions we will be interested in, it will lead to theories that enjoy various types of asymptotic symmetry, making them highly non-generic in the sense discussed in Sec. \ref{sec2}. 

\subsection*{Perturbations}

As a next step, we check whether the cosmological solutions obtained through the above procedure are stable. This can be done with the help of the analysis spelled out in Appendix \ref{appA}. In the unitary gauge, defined by the absence of $\pi$ - fluctuations, $\pi(x,t)=\pi_0(t)$, the only scalar degree of freedom present in the theory is captured by the standard curvature perturbation of equal-density hypersurfaces $\zeta$, that enters into the perturbed spatial metric in the following way  
\beq
\label{gij}
g_{ij}=a(t)^2 (1+2\zeta)\delta_{ij}~.
\eeq 
The curvature perturbation is an exactly massless field, which directly follows from the fact that $\zeta =const$ should be a legitimate solution, since $g_{ij}$ in this case is obtained from the unperturbed FRW metric by a mere constant rescaling of spatial coordinates (this, of course, is also the origin of conservation of $\zeta$ at super-horizon distances).  

Having the background quantities at hand, one can readily derive the quadratic $\zeta$ action following the standard procedure \cite{Maldacena:2002vr} 
\beq
\label{quadact}
S_\zeta=\int d^4x~ a^3~\bigg[A(t)~\dot\zeta^2-B(t)~\frac{1}{a^2}\(\vec{\nabla}\zeta\)^2-C(t)~\frac{1}{a^4}\(\vec{\nabla}^2\zeta\)^2   \bigg]~.
\eeq
The kinetic coefficients $A$ and $B$ are found to be \cite{Creminelli:2006xe,Creminelli:2010ba} 
\bea
\label{A}
A(t) &=&\frac{\mpl^2 (-4 \mpl^4 \dot H-12\mpl^2 H \hat M^3+3\hat M^6+2\mpl^2 M^4)}{(2\mpl^2H-\hat M^3)^2}~,\\
\label{B}
B(t)&=&\frac{\mpl^2 \(-4 \mpl^4 \dot H+2\mpl^2 H \hat M^3-\hat M^6+2\mpl^2\p_t\hat M^3\)}{(2\mpl^2H-\hat M^3)^2}~,
\ea
while $C(t)=0$ for our `classical' action \eqref{ggg} (it will be nonzero once we include higher-order terms in the effective theory in Sec. \ref{hdim}).
Explicit expressions for the time-dependent coefficients $\hat M^3$ and $M^4$ are given in Eq. \eqref{M's}. 
 Apart from other background quantities, these explicitly depend on the function $\mathcal{F}_2(\pi_0)$. Using the expression \eqref{f2} for the latter, one finds
\ba
\label{Anec}
A&=&3 \mpl^2 ~\frac {36 \mpl^4 H_0^4 H^2+9\mpl^4 H_0^4\dot H-18 \mpl^2 f^2 H_0^2H \dot\pi ^3 -6\mpl^2 f^2 H_0^2 \dot\pi ^2\ddot\pi+4 f^4 \dot\pi^6}{(3\mpl^2 H_0^2 H-2f^2 \dot\pi^3)^2},~~~~~~~~\\ 
\label{Bnec}
B&=&\frac{-9\mpl^6 H_0^4 \dot H+6\mpl^4 f^2H_0^2H\dot\pi^3+18\mpl^4 f^2  H_0^2 \dot \pi^2\ddot\pi  -4\mpl^2 f^4\dot\pi^6}{(3\mpl^2 H_0^2 H-2f^2 \dot\pi^3)^2},
\ea
while the speed of sound for short wavelength scalar perturbations is given by $c_s^2=A/B$.
Positive $A$ and $B$ throughout the entire course of cosmological evolution guarantee the absence of ghost and gradient instabilities respectively. 

As a quick check, one can apply the above piece of formalism to galilean genesis \cite{Creminelli:2010ba}.
Plugging the scalar and Hubble profiles, \eqref{ggsol} and \eqref{ggsol1} into the expressions for the curvature perturbation's kinetic coefficients \eqref{Anec} and \eqref{Bnec}, one obtains the following values for the latter quantities to the leading order in $\mpl$:
\beq
A(t)=B(t)=\frac{9\mpl^4H_0^2}{f^2} ~t^2~.
\eeq
This precisely agrees with the expressions found in \cite{Creminelli:2010ba}.

\section{Extended genesis: analytic solutions}
\label{analytic}

While the recipe, spelled out in the previous section formally allows to construct theories admitting essentially arbitrary cosmological solutions, most of these fail to be physically meaningful in one way or another. A generic such solution will lead to either ghost or gradient instability at the level of small perturbations; moreover, most of the resulting theories will be free from symmetries -- even the asymptotic ones, casting shadow on quantum robustness of the whole picture. Nevertheless, we will show in this section that a class of theories exists, that admit completely stable cosmological solutions interpolating between a low/zero curvature maximally symmetric spacetime in the far past and a larger curvature inflationary dS spacetime in the future -- with a strong/moderate violation of the null energy condition in between. Importantly, we will see that asymptotically these theories enjoy symmetries of the kind described in Sec. \ref{sec2}.

Let us work in a coordinate system such that time runs from $t =-\infty$ towards $t=0$ over the cosmological phase of interest. At (or shortly after) $t=0$, the system is assumed to reheat, or exit the given phase otherwise. Inspired by the early-time galilean genesis asymptotics \eqref{ggsol} and \eqref{ggsol1}, we will adopt the following ansatz for the Hubble rate
\beq
\label{ansatz}
H=\lambda +\beta ~\frac{f^2}{\mpl^2 H_0^2} ~\dot\pi_0^3~,
\eeq
where $\lambda$ and $\beta$ are free parameters (of mass dimension one and zero respectively) of the theory, giving rise to the solution of interest.
For the scalar, we will assume the ansatz of the following form (which is again motivated by the genesis solution)
\beq
\label{ansatz1}
e^{\pi_0}=\frac{1}{H_0} \frac{1}{t_*-t}~.
\eeq
Here, $t_*>0$ is yet another free parameter with mass dimension minus one. While resembling GG at early times (and for sufficiently small $\lambda$), \eqref{ansatz} and \eqref{ansatz1} describe a cosmology regularized towards $t\to 0^-$, so that none of the invariants in the theory grow unbounded over the entire interval $t\in [-\infty,~0]$. 
Galilean genesis is recovered at all times for the particular values of the parameters $\lambda=0$, $\beta=1/3$ and $t_*=0$. For $\lambda\neq 0$ on the other hand, there is a crucial difference: rather than from flat, Minkowski spacetime, the system starts out evolving from de Sitter space with the curvature set by the parameter $\lambda$. 

In order for the universe to be described by inflationary de Sitter geometry at $t\to 0^-$, the parameters of the theory should satisfy certain constraints.   
One such constraint arises from requiring the Hubble rate not to vary considerably over a single e-fold at $|t|\ll t_*$. The necessary condition for that is: 
\beq
\label{dscond}
1\gg \varepsilon \equiv \frac{\dot H}{H^2}\bigg |_{t\to 0}\sim 
\begin{cases}
\frac{\mpl^2 H_0^2}{\beta f^2} ~t_*^2~, &\text{if} ~~\lambda \ll \beta~\frac{ f^2}{\mpl^2 H_0^2}~\frac{1}{ t_*^3} \\
\frac{\beta}{\lambda^2}~ \frac{f^2}{\mpl^2 H_0^2}~\frac{1}{ t_*^4}~, &\text{if} ~~ \lambda \gg \beta~\frac{ f^2}{\mpl^2 H_0^2}~\frac{1}{ t_*^3}~.
\end{cases}
\eeq
Not surprisingly, this condition is equivalent to the one constraining $\dot \pi$ to be quasi-constant at late times: 
\beq
\frac{1}{H}\frac{d}{dt} \ln \dot\pi_0\ll 1 ~.
\eeq
This shows that $\pi$ can indeed be approximated by a linear profile towards $t\to 0^-$, leading to galileon inflation discussed in Sec. \ref{sec2}.

In the rest of this section we will study various interesting regions in the six-dimensional space spanned by the free parameters $\( \mpl, f,H_0,\lambda,\beta, t_* \)$ of the theory.

\subsection{$\lambda =0$} 

We begin with the case that, in the asymptotic past, the system starts out evolving from flat spacetime. This happens for $\lambda=0$. As a quick consistency check, one can \textit{derive} the conformally invariant GG lagrangian \eqref{gg} from our ansatz for the extended genesis cosmology, following the inverse construction of the previous section. Indeed, plugging \eqref{ansatz} and \eqref{ansatz1} (with $\beta=1/3$) into the expressions for $\mathcal{F}$-functions, \eqref{f1} and \eqref{f2}, we find at the leading order in $1/\mpl^2$ (and at times $|t|\gg t_*$)
\beq
\mathcal{F}_1=\frac{1}{H_0^2 t^2}=e^{2\pi}, \qquad \mathcal{F}_2=1~.
\eeq
This precisely corresponds to the conformal galileon. For values of $\beta$ other than $1/3$, on the other hand, our ansatz describes subluminal versions of GG \cite{Creminelli:2012my} at $|t|\gg t_*$.

Concentrating on the full solution, including times $|t|\leq t_*$, stability of the system requires that the kinetic coefficients in \eqref{quadact} are positive at all times. For $\lambda=0$, they are given as follows 
\ba
\label{atilde}
\frac{3 (2\mpl^2 H-\hat M^3)^2 H_0^4}{4\mpl^2}~A&=&2(4+15 x +18x^2) f^4 \dot\pi^6+3(4+9x)\mpl^2 f^2 H_0^2\dot\pi^2\ddot\pi~,~~~~~\\
\label{btilde}
\frac{3 (2\mpl^2 H-\hat M^3)^2 H_0^4}{4\mpl^2}~B&=& \(2 x f^2\dot\pi^4-9x \mpl^2 H_0^2\ddot\pi\)f^2 \dot\pi^2~,
\ea
where we have defined $x=\beta-2/3$ for further convenience. As an immediate observation, we note that $A$ is manifestly positive for positive $x$ (both $\dot\pi$ and $\ddot\pi$ are positive at all times for our ansatz), while $B$ does not have a definite sign. For the special  case that the parameter $x$ is small however, $B$ can be made arbitrarily small, compared to $A$, implying a vanishing speed of sound for $\zeta$. This is similar to what happens in ghost condensation, where the absence of gradient instability is determined by higher-order operators in the effective theory.

It is straightforward to see that $B$ cannot be positive over the entire temporal interval of interest -- at least for our ansatz \eqref{ansatz}. Indeed, we are interested in solutions, that start in galilean genesis at $t\to -\infty$ and end up in the inflationary phase at $t\to 0^-$. 
As shown in the previous section, the latter phase requires $\dot\pi$ to be practically constant, meaning that the second term in the parentheses on the r.h.s. of \eqref{btilde} should be negligible compared to the first one at late times. Positivity of $B$ at late times then requires $x>0$. On the other hand, galilean genesis corresponds to the second term prevailing at sufficiently early times, since $\ddot \pi\sim 1/t^2$ decreases parametrically slower than $\dot\pi^4\sim 1/t^4$ at large and negative $t$. For $x>0$ however, this would lead to gradient instability at early times. In contrast, in the opposite case of $x<0$, one would recover gradient instability at late times, while the early-time genesis phase would be completely stable. One is therefore led to conclude that gradient instability is unavoidable for the given choice of the ansatz \eqref{ansatz} in the $\lambda=0$ case -- at least at the leading order in derivative expansion.

Concentrating on negative $x$ (so that the genesis phase is stable), the time at which gradient instability occurs (i.e. when $B$ flips sign) is of order $|\tau|\sim f/(\mpl H_0)$. The slow-roll parameter at that time can be readily estimated, $\varepsilon\sim \mpl^2 H_0^2 \tau^2/f^2\sim 1$, see Eq. \eqref{dscond}. This means that the gradient instability for $\lambda=0$ solutions necessarily kicks in before the onset of the de Sitter regime, explaining the pattern we have found via numerical analysis in Sec. \ref{sec2} (see also Appendix \ref{num}).

We end the present subsection with a couple of consistency checks for our calculations. First, we note that for $x=-1/3$ corresponding to galilean genesis, one recovers an exactly luminal scalar mode, $c_s^2=\tilde B/\tilde A=1$ at early times. Moreover, as stressed several times above, the late-time de Sitter phase should correspond to an enhanced shift symmetry on $\pi$. That this is indeed the case is the result of quasi-constancy of the $\mathcal{F}$ functions
\beq
\frac{1}{H} ~\frac{d}{dt}\ln \mathcal{F}_{1,2} \ll 1~.
\eeq
which, as can be straightforwardly verified, directly follows from (the $\lambda=0$ version of) Eq. \eqref{dscond} -- the condition for the universe to be described by de Sitter geometry at $|t|\ll t_*$.

\subsection{$\lambda \neq 0$} 

We now turn to the case that in the asymptotic past the universe starts out evolving from de Sitter space, rather than Minkowski, $\lambda\neq 0$. The curvature of the initial state is of order $R\sim \lambda^2$ and is a free parameter of the theory; if its value is strictly zero, we have seen that the resultant cosmological solution suffers from a gradient instability before the onset of de Sitter regime for much of the parameter space -- at least if one ignores higher-order operators in the effective theory. However, for non-zero $\lambda$, as we will now demonstrate, gradient instabilities can be avoided even in the 'classical' theory, that is without invoking higher-derivative terms in the EFT for perturbations.  

The kinetic coefficients \eqref{Anec} and \eqref{Bnec}, evaluated on the given ansatz are:
\ba
A&=&\frac{\mpl^2}{3}~ \frac{36  \mpl^4 H_0^4 \lambda^2\tau^6+3 \mpl^2 f^2 H_0^2 [(-(10+24 x)\lambda\tau +4+9x )]\tau ^2+f^4 (8+30x+36 x^2)}{(f^2 x-\mpl^2 H_0^2\lambda \tau^3)^2}~,\nn \\
B&=&\frac{\mpl^2}{3}~ \frac{\mpl^2 f^2 H_0^2 (-2\lambda\tau-9x)\tau^2 + 2 f^4 x}{(f^2 x-\mpl^2 H_0^2\lambda \tau^3)^2}~\nn ,
\ea
where we have defined $\tau\equiv t-t_* \leq -t_*$ . An important observation that we will use in what follows is that for positive $x$, and for $\bar \varepsilon \equiv \lambda t_*>9 x/2$, both $A$ and $B$ are manifestly positive (and finite) \textit{at all times}, as can be readily verified by inspecting the above expressions. 

Given that a strictly vanishing $\lambda$ is not allowed by stability, how small can it be? The smallness of the initial curvature can be conveniently characterized by 
\beq
\label{hsep}
\frac{H(t=0)}{H(t=-\infty)}=\(\lambda~\frac{\mpl^2 H_0^2 t_*^3}{\beta f^2}\)^{-1}\sim \frac{1}{\varepsilon \bar\varepsilon}~.
\eeq
Note that, while $\varepsilon\ll 1$ is required by the late-time de Sitter space, $\bar \varepsilon$ is in principle an unconstrained parameter of the theory. 

To summarize, choosing $x< 2\bar \varepsilon/9$, one can arrange for a manifestly stable cosmological solution, interpolating between two de Sitter spacetimes with an arbitrary ratio of the corresponding asymptotic curvatures. Moreover, the larger is the separation between the asymptotic Hubble rates \eqref{hsep}, the smaller is the deviation of the late-time geometry from perfect de Sitter space. The speed of sound of the curvature perturbation at $t=0$ can be readily evaluated from the above expressions for the kinetic coefficients
\beq
\label{cssq1}
c_s^2(t=0)=\frac{x (2-9 \varepsilon)+2\varepsilon \bar\varepsilon}{8+30 x+36 x^2+\mathcal{O}(\varepsilon)}~.
\eeq 
Note that the asymptotic $c_s^2$ is finite. For $\varepsilon=0$, its magnitude is bounded from above by $c_s^2<0.031$, which can be found by maximizing the expression \eqref{cssq1} for the squared speed of sound\footnote{Cf. the analytic bound on the scalar speed of sound $c_s^2<0.031$ in galileon inflation, quotted in \cite{Kobayashi:2010cm}.}.

Let us for simplicity set $x= 0$ from now on. One distinct property of our ansatz is that the coefficient $A$, having a contribution constant in time, becomes parametrically greater than $B$ at $|t|\gg t_0$, as $B\sim -1/t^3$ at large and negative $t$. This means that the speed of sound of the curvature perturbation tends to zero at early times. 

What is the theory describing the asymptotic past of the background solutions at hand? To answer this question, we evaluate the $\mathcal{F}$ functions from our deformed galileon action \eqref{ggg}. At the leading order in $1/t$, one finds
\ba
\mathcal{F}_1&=&6~\frac{\mpl^2 \lambda^2}{f^2 H_0^2}~ (H_0 t)^2=6~\frac{\mpl^2 \lambda^2}{f^2 H_0^2}~e^{-2\pi}~,\nn \\
\mathcal{F}_2&=&9~\frac{\mpl^2 \lambda^2}{f^2 H_0^2}~ (H_0 t)^4=9~\frac{\mpl^2 \lambda^2}{f^2 H_0^2}~e^{-4\pi}~,\nn
\ea 
which implies the following form of the scalar action 
\beq
\label{earlydslag}
\mathcal{S}^{\text{early}}_\pi=\int d^4 x ~\sqrt{-g} ~\bigg[ 6~\frac{\mpl^2 \lambda^2}{ H_0^2}~e^{-2\pi} (\p\pi)^2+\frac{2}{3}~\frac{f^2}{H_0^2} (\p\pi)^2\Box\pi +3~\frac{\mpl^2 \lambda^2}{H_0^4}~e^{-4\pi}  (\p\pi)^4    \bigg ]~.\nn
\eeq 
In the regime of interest, $e^\pi\sim1/t$ and the first and the third terms in the parentheses are constant, while the second (the cubic galileon) goes as $\sim 1/t^3$ and is thus completely irrelevant in the asymptotic past\footnote{The latter estimate comes from the $H \dot \pi^3 $ piece, coming from the expansion of the covariant derivative on a de Sitter background.}. 
\begin{figure}
\centering
\includegraphics[width=.46\textwidth]{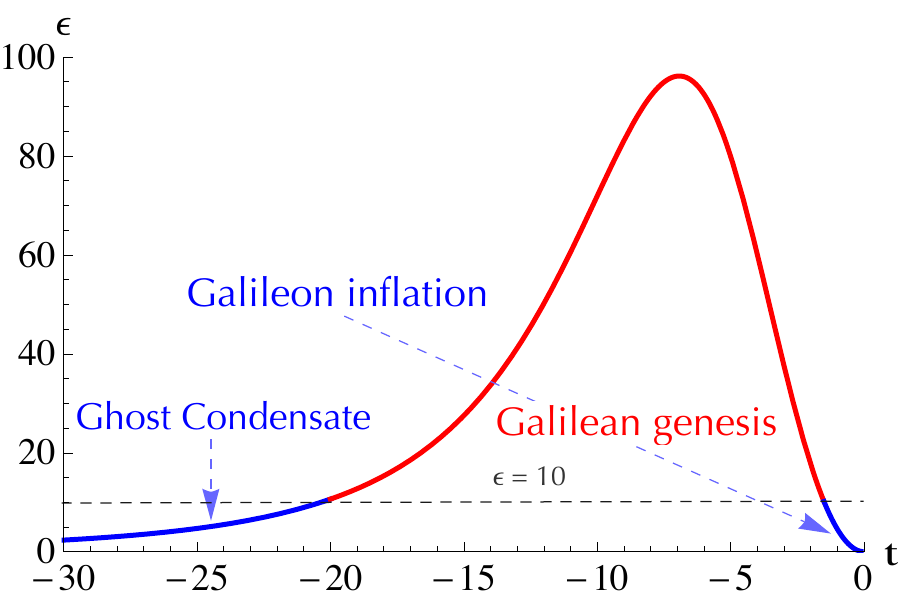} \quad
\includegraphics[width=.46\textwidth]{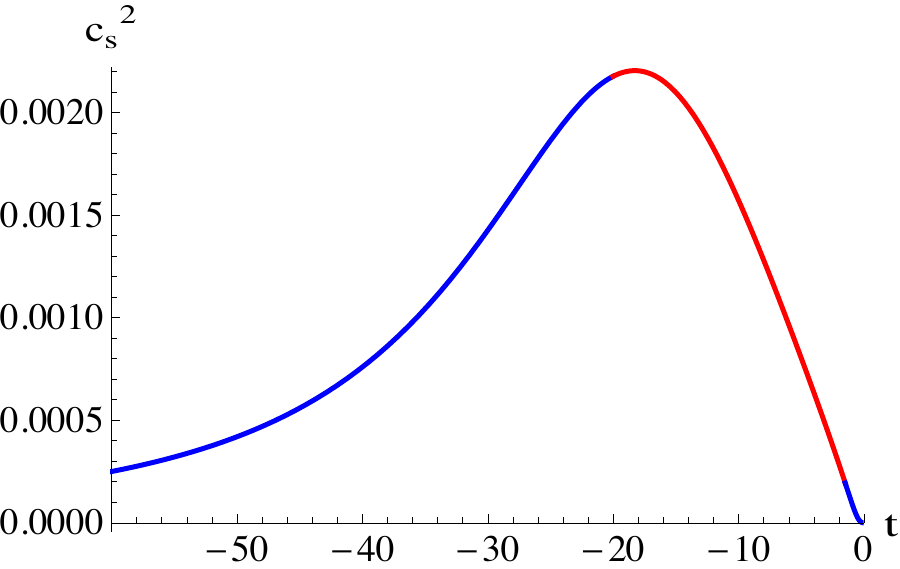}
\caption{The 'slow-roll' parameter $\varepsilon$ (left) and the speed of sound of the curvature perturbation $c_s^2$ (right) as functions of time on the solution \eqref{ansatz}, \eqref{ansatz1}. The scales $f$ and $\mpl$ have been assumed equal, while the rest of the parameters have been chosen to be: $H_0=1, ~\lambda=10^{-3},~ t_*=10^{-2},~ x=0$. The two colors correspond to $\varepsilon <10$ (blue) and $\varepsilon>10$ (red).}
\label{figurethree}
\end{figure}
Once the cubic galileon is neglected however, the theory acquires a global symmetry. To see it, it is useful to define a new field $\chi=e^{-\pi}$, in terms of which the two relevant operators are simply $(\p\chi)^2$ and $(\p\chi)^4$, and the new symmetry is immediately identified as invariance under constant shifts $\chi\to \chi+c$ (while in terms of $\pi$ this symmetry looks more complicated: $\pi\to -\ln \(e^{-\pi}+c\)$). This shows, that the early-time $\lambda\neq 0$ cosmology is effectively described by a ghost condensate - type theory, albeit written in obscure variables (and hence the vanishing speed of sound)! Needless to say, the emergent global symmetry comes hand-in-hand with all the attractive properties, classical or quantum, characteristic of ghost condensation\footnote{That the given solution indeed describes ghost condensation can also be seen from the fact that $\chi$ acquires a linear profile, $\chi=-H_0 t$, just as the ghost field does on self-accelerated backgrounds.}, see \cite{ArkaniHamed:2003uy,ArkaniHamed:2003uz}. 

To get a more quantitative perspective on the above discussion, let us consider the solutions \eqref{ansatz} and \eqref{ansatz1} for a specific set of available parameters. As an immediate observation, we note that the Hubble rate does not depend on the magnitude of $f$ and $\mpl$ separately (as far as external matter or bare cosmological constant are not introduced into the system) -- physical quantities are only sensitive to the ratio of the two scales. As a result, one can arbitrarily set the physical units for any one quantity at any one instant of time. For example, the Hubble scale at time $t=0$ can be freely chosen to be $H(0)=10^{14} ~GeV$ in some putative system of units where $H_0\equiv 1$. With this in mind, we set $f=\mpl$, and consider the following values for the rest of the parameters: $H_0=1, ~\lambda=10^{-3},~ t_*=10^{-2},~ x=0$, satisfying (the first case of) the late-time dS condition, Eq. \eqref{dscond}.

The time-dependence of the `slow roll' parameter $\varepsilon=\dot H/H^2$ (left) and the speed of sound of the curvature perturbation $c_s^2$ (right) for the above choice of the theory parameters is shown in Fig. \ref{figurethree}. From how $\varepsilon$ depends on time, one can distinguish three stages of evolution, according to whether the system violates the NEC strongly (red), or weakly (blue). The universe starts out in de Sitter space ($\varepsilon \simeq 0$) with tiny curvature $\sim \lambda^2$, the Hubble rate as well as the slow-roll parameter $\varepsilon$ gradually increasing with time. When $\varepsilon \simeq 10$, it enters into the galilean genesis phase with strong violation of the null energy condition.  Peaking at $\varepsilon \sim 10^2$ at intermediate times, NEC-violation weakens down back to $\varepsilon \simeq 10$ at $t\simeq -1.5$ (signalling the beginning of the third, galileon inflation stage), $\varepsilon$ decreasing to sub per-cent values shortly afterwards (the final phase of the system corresponds to the blue ends of the curves near $t\to 0$ in Fig. \ref{figurethree}). 

While the concluding, inflationary de Sitter phase seems rather short in its extension in time, the large magnitude (in units of $H_0$) of the expansion rate at those times allows it to accomodate a large number of e-folds. Indeed, from $t=-0.1$ ($\varepsilon \simeq 5\cdot 10^{-2}$) up until $t=0$ ($\varepsilon \simeq 5\cdot 10^{-4}$), the number of times the scale factor doubles can be easily estimated
\beq
N_e=\int\limits_{-0.1}\limits^{0} H dt\simeq 3300~,
\eeq
showing that the de Sitter phase towards the end of the temporal interval of interest is in fact very extended. Furthermore, the Hubble parameter at $t=0$ is $H(0)\sim 10^6$, implying a huge ratio of de Sitter expansion rates in the asymptotic future and the asymptotic past
\beq
\frac{H(0)}{H(-\infty)}\sim 10^9~.
\eeq

The right panel of Fig. \ref{figurethree} shows the evolution of the scalar speed of sound. As remarked above, $c_s^2$ starts evolving from nearly zero value at early times, as required by ghost condensate-type cosmologies. Peaking at $c_s^2\simeq 2\cdot 10^{-3}$ during the genesis stage, it drops down again towards late-time galileon inflation.

While ghost condensation, described by a $P(X)$-type theory implies vanishing  speed of sound of the scalar perturbation at the leading order \cite{ArkaniHamed:2003uy}, galileon inflation (described by a $P(X)$ lagrangian plus one or more galileon terms) does not necessarily lead to $c_s^2=0$ although, as discussed before, there is an upper bound $c_s^2\leq 0.031$ in the latter class of models with a single cubic galileon \cite{Kobayashi:2010cm}. 
Our solutions however qualitatively (and crucially) differ from 'tilted' ghost condensate with NEC violation considered in \cite{Creminelli:2006xe} in that the speed of sound, although small, is strictly \textit{positive} at all times. The latter is not true for pure $P(X)$ theories: violation of the null energy condition unambiguously implies gradient instabilities at the leading order in the ghost condensate \cite{Hsu:2004vr,Dubovsky:2005xd}.

At early times, the tiny speed of sound of the scalar mode suggests that higher-order operators in the effective theory for perturbations \cite{Creminelli:2006xe,Cheung:2007st} could be qualitatively affecting the dynamics of the system. Moreover, depending on the nature of the UV completion, higher-derivative terms could also play a role in the intermediate, galilean genesis phase. In order to estimate these effects, we turn to exploring the structure of the next-to-leading-order action in the EFT formalism in the following section.

\section{Beyond the leading order}
\label{hdim}

The tiny asymptotic scalar speed of sound found for the EG solutions motivates to go beyond the leading order in the EFT for perturbations to assess the role of higher-derivative operators in stability of the system. The generic action for metric fluctuations on a FRW background driven by a single `clock' has the following form (excluding the Einstein-Hilbert part) \cite{Creminelli:2006xe,Cheung:2007st}
\ba
\label{s_pi}
S_\pi&=&\int d^4x~\sqrt{g_3}N\bigg[-\mpl^2 \dot H \frac{1}{N^2}-\mpl^2(3 H^2+\dot H)\nn \\
&+&\frac{1}{2} M^4(t) (\delta N)^2-\hat M_3^3(t)\delta E^i_{~i}\delta N -\frac{\bar M'(t)^2}{2}\delta E^{ij} \delta E_{ij}-\frac{\bar M(t)^2}{2}\delta E^{i~2}_{~i}+\dots  \bigg],
\ea
where $g_3,~N$ and $N_i$ are the standard ADM variables \cite{Arnowitt:1962hi}, while $E_{ij}$ is related to the extrinsic curvature of equal-time hypersurfaces, see Appendix \ref{appA} for a detailed discussion. Furthermore, $\delta N$ and $\delta E_{ij}$ denote perturbations of the corresponding quantities over their background values. The `classical' theory \eqref{ggg} generates only the first two terms on the second line of \eqref{s_pi}, and all of the above analysis has assumed vanishing $\bar M$ and $\bar M'$ (as well as yet higher-derivative operators, implied by the ellipses). In practice, the latter coefficients are expected to be present, although suppressed in derivative expansion. 

In what follows, we assume nonzero $\bar M^2$ and $\bar M'^2$ in computing the quadratic action for $\zeta$ on extended genesis backgrounds \footnote{Both $\bar M^2$ and $\bar M'^2$ can in principle have either sign. The notation used for these coefficients only serves to emphasize their mass dimension. }. The results, given in \eqref{A'}-\eqref{C'} of Appendix \ref{appA}, are rather tedious and reluctant to simple analysis in their exact form. To simplify life, we will expand all relevant quantities to linear order in  $\bar M^2$ and $\bar M'^2$, assuming these are small in the sense that higher order terms in the expansion give subleading corrections -- something we will justify \textit{a posteriori}. The procedure yields the following expressions for the kinetic coefficients\footnote{The signs are defined so that all kinetic coefficients have to be positive for complete stability (stability at all wavelengths) of the corresponding background.} $A$, $B$ and $C$ 
on backgrounds corresponding to the second, $\lambda\neq 0$ case of the previous section 
\ba
A&=&\frac{2}{3\mpl^2 H_0^4\lambda^2\tau^6}~\big(18 \mpl^4 H_0^4\lambda^2 \tau^6-3\mpl^2f^2H_0^2(5\lambda \tau-2)\tau^2+4 f^4\big )+p_1\bar M^2+p_2\bar M'^2,\nn \\
B&=&-\frac{2}{3}\frac{f^2}{H_0^2\lambda}\frac{1}{\tau^3}+p_3\bar M^2+p_4\bar M'^2+q_3\p_t(\bar M^2)+q_4\p_t(\bar M'^2),\nn \\
C&=&\frac{\bar M^2+\bar M'^2}{2\lambda^2},\nn
\ea
where we have defined $\tau \equiv t-t_*<0$ and introduced auxiliary coefficients $p_i$ and $q_i$, given as follows
\ba
p_1&=&-\frac{1}{18 \mpl^8 H_0^8\lambda^4\tau^{12}}~\big (27 \mpl^4 H_0^4\lambda^2\tau^6-6\mpl^2 f^2 H_0^2(5\lambda\tau-2)\tau^2+8 f^4\big )^2, \nn \\
p_2&=&-\frac{1}{18 \mpl^8 H_0^8\lambda^4\tau^{12}}~\bigg [ 1107 \mpl^8 H_0^8\lambda^4\tau^{12}-1980\mpl^6 f^2H_0^6\lambda^3\tau^9 +792\mpl^6 f^2H_0^6\lambda^2\tau^8\nn \\&+&1428\mpl^4 f^4 H_0^4\lambda^2\tau^6-720 \mpl^4 f^4 H_0^4\lambda \tau^5+144\mpl^4 f^4 H_0^4\tau^4-480\mpl^2 f^6 H_0^2\lambda\tau^3\nn \\
&+&192 \mpl^2 f^6 H_0^2 \tau^2+64 f^8 \bigg ],\nn\\
p_3&=&-\frac{1}{18\mpl^6 H_0^6 \lambda^3\tau^9}~\bigg[81\mpl^6 H_0^6\lambda^3 \tau^9-18 \mpl^4 f^2 H_0^4 (8\lambda^2\tau^2-17\lambda\tau+8)\tau^4\nn \\ &+&84 \mpl^2 f^4 H_0^2 (\lambda \tau-2)\tau^2-16 f^6 \bigg],\nn \\
p_4&=&-\frac{1}{18\mpl^6 H_0^6 \lambda^3\tau^9}~\bigg[99\mpl^6 H_0^6\lambda^3 \tau^9-6\mpl^4 f^2 H_0^4 (26\lambda^2\tau^2-51\lambda\tau+24)\tau^4\nn \\ &+&84 \mpl^2 f^4 H_0^2 (\lambda \tau-2)\tau^2-16 f^6 \bigg],\nn \\
q_3&=&-\frac{1}{6 \mpl^4 H_0^4\lambda^3\tau^{6}}~\big[ 27\mpl^4 H_0^4 \lambda^2\tau^6-6\mpl^2 f^2 H_0^2(5\lambda \tau-2)\tau^2+8 f^4 \big],\nn \\
q_4&=&-\frac{1}{6 \mpl^4 H_0^4\lambda^3\tau^{6}}~\big[ 33\mpl^4 H_0^4 \lambda^2\tau^6-6\mpl^2 f^2 H_0^2(5\lambda \tau-2)\tau^2+8 f^4 \big]~.\nn
\ea
An immediate and important observation is that all of the coefficients $p_i$ and $q_i$ are sign-definite (negative) \textit{at all times}. Moreover, since different linear combinations of $\bar M^2$ and $\bar M'^2$ enter into $B$ and $C$, nothing prevents us from choosing the former pair of EFT coefficients such that both $B$ and $C$ are positive - thus avoiding any instability over the entire cosmological period of interest!

Furthermore, we have found in the previous section that the speed of sound of the scalar mode tends to zero ($c_s^2\to 0^+$) in the asymptotic past for the backgrounds corresponding to EG. This suggests that $t\to -\infty$ is precisely the regime where higher-order corrections in the EFT for perturbations could play an important role. In fact, in the case that $\bar M, \bar M'$ fall off slower than $1/t^3$ at early times, higher-order effects give contributions that dominate over the leading-order piece in the coefficient $B$ at early times\footnote{This seemingly casts shadow on the very meaning of our expansion in small $\bar M, \bar M'$; fortunately, a closer inspection of \eqref{A'}-\eqref{C'} shows that the expansion parameters at $t\to -\infty$ are in fact $\bar M^2/\mpl^2$ and $\bar M'^2/\mpl^2$ -- meaning that next-order corrections in the series indeed give subleading effects.}. Focusing on the $t\to-\infty$ ghost condensate regime (and neglecting time derivatives of $\bar M, \bar M'$ for simplicity), we find
\ba
\label{App}
A&=&12 \mpl^2  +\mathcal{O}\(\bar M^2,\bar M'^2\)~, \\
\label{Bpp}
B&=&-\frac{2}{3}\frac{f^2}{H_0^2\lambda}\frac{1}{t^3}-\frac{9}{2} \bar M^2-\frac{11}{2} \bar M'^2+\mathcal{O}\(\frac{\bar M^4}{\mpl^2},\frac{\bar M'^4}{\mpl^2}\)~, \\
\label{Cpp}
C&=&\frac{\bar M^2+\bar M'^2}{2 \lambda^2}+\mathcal{O}\(\frac{\bar M^4}{\mpl^2\lambda^2},\frac{\bar M'^4}{\mpl^2\lambda^2}\) ~.
\ea
Again, since $B$ and $C$ involve different linear combinations of $\bar M^2$ and $\bar M'^2$, one can freely choose the values for the latter two coefficients, such that the theory is free from any instability\footnote{Note that there is in fact even more freedom: one could always make the coefficient $C$ positive by adding a term of the form $\sqrt{g_3} N R_3^{~2}$ (which does not affect the quadratic action for tensor perturbations) to \eqref{s_pi}, see the discussion in Appendix \ref{num}.}. 

Moreover, in the case that $\bar M, \bar M'$ drop off slower than $1/t^3$ for large and negative times, the asymptotic speed of sound of the scalar perturbation is set by the ratio 
$$c_s^2 = \frac{|9 \bar M^2+11\bar M'^2|}{24 \mpl^2}~,$$ and is not necessarily infinitesimally close to zero if at least one of the two EFT coefficients $\bar M$ and $\bar M'$ tends to a constant at early times\footnote{Note, that the squared speed of sound also sets the magnitude for the expansion parameter in \eqref{App}-\eqref{Cpp}.}.

To summarize, we have found that beyond-the-leading-order structure of the effective theory for perturbations does possess enough freedom to allow to 
cure (weak) classical gradient instability.
Whenever the speed of sound of the scalar mode vanishes at the leading order on the other hand, higher-order effects can push the corresponding solution into a completely stable direction.

\section{Conclusions and future directions}
\label{concl}

Despite the extremely compelling picture of the early cosmology that standard slow-roll inflation provides us with, it is still fair to say that it is not the only possible one. The question of how far alternative scenarios can go in adequately describing the observed universe has been a strong motivation for expanding the theory space in non-standard directions. Perhaps the most dramatic departure from the inflationary paradigm corresponds to theories that violate the null energy condition, thereby allowing for a qualitatively different evolution of the early universe that, among other interesting features, is capable of smoothing out the Big-Bang singularity. That this can happen without instabilities for a universe starting out from the flat, Minkowski spacetime has been shown in Refs. \cite{Creminelli:2006xe,Creminelli:2010ba,Hinterbichler:2012fr}. 

A common feature of alternatives to inflation based on NEC violation is that they usually predict a strongly blue-tilted and unobservable (at least in the CMB experiments) spectrum of tensor perturbations. The ultimate reason for this lies in the fact that the phenomenologically interesting phase of cosmological evolution happens on quasi-flat backgrounds. Would then a detection of primordial $B$ modes in CMB polarization conclusively rule out these theories? 

In this paper we have argued that the answer to this question is negative.   
We have constructed explicit theories that lead to an early universe cosmology interpolating between a small/zero curvature maximally symmetric (dS or Minkowski) spacetime in the far past and an inflationary de Sitter spacetime, capable of generating a scale-invariant tensor spectrum of significant amplitude in the asymptotic future; this is possible because, at intermediate times, the system can \textit{strongly} violate the null energy condition ($\dot H\gg H^2$) as it happens in genesis models -- all without developing any kind of instability. The corresponding backgrounds can be viewed as a regularized extension of galilean genesis -- one for which none of the physical quantities grow beyond the cutoff scale. Alternatively, one can view them as a certain `UV' (or, to be more precise, an early-time) -complete realization of inflation, that leads to a (almost) flat pre-inflationary universe.

Being deformations of the conformal galileon, the theories constructed above enjoy non-linearly realized emergent symmetries at both the early- and the late-time asymptotics. It is in fact precisely the nature of the asymptotics that determines the qualitative picture of the cosmological solutions of interest: these are described by quantum-mechanically stable, robust theories based solely on symmetry principles. An alternative, and very interesting realization of genesis cosmologies occurs in the context of the Dirac-Born-Infeld (DBI) models \cite{Hinterbichler:2012fr, Hinterbichler:2012yn}. Needless to say, it would be interesting to see how our construction carries over to theories enjoying asymptotic DBI-like symmetries. 

At this stage, the presented models are not intended as fully realistic, however upon slight adjustment they should become capable of facing observational challenges. While realistic model-building is beyond the scope of this paper, we briefly list the phenomenological questions that remain to be addressed. Above all, a mechanism for exiting the inflationary de Sitter regime/reheating has to be specified\footnote{This can be done \textit{e.g.}  by giving a step function-like potential to the scalar as in ghost inflation -- a mechanism that can be implemented in a technically natural way since the asymptotic shift symmetry breaking becomes localized in field space in this case, see, \textit{e.g.} \cite{ArkaniHamed:2003uz}.}. Another question is that of the observed negative scalar tilt, which is not characteristic of NEC- violating inflationary theories with the inflaton being the field responsible for adiabatic perturbations. The negative tilt of density perturbations can arise from small shift-symmetry breaking effects (necessary to end the de Sitter phase), or through the standard mechanisms such as curvaton \cite{Enqvist:2001zp,Lyth:2001nq} or inhomogeneous reheating \cite{Dvali:2003em}. Furthermore, we have seen that on extended genesis backgrounds, the scalar perturbations are characterized by a relatively small speed of sound. Quite generally, small scalar speed of sound translates into large equilateral non-gaussianity \cite{Cheung:2007st} -- a result that follows solely from the requirement of nonlinearly realizing the broken time translation invariance\footnote{In addition, the small scalar speed of sound leads to an interesting effect, whereby the scalar perturbations of a given comoving wavelength freeze out earlier than the tensor modes leading to an enhancement of the tensor-to-scalar ratio, see \textit{e.g.} the recent discussion of Ref. \cite{Baumann:2014cja}. In our context, this could lead to a striking possibility of scalars freezing out in the genesis phase, while the tensors -- in the inflationary one. }. These, among other phenomenological aspects, will be discussed elsewhere.

Putting aside phenomenology, our models serve as a proof of principle for the possibility to smoothly and stably connect an inflationary quasi-de Sitter universe to a much lower, or even zero-curvature, maximally symmetric spacetime in the asymptotic past -- all without exiting the regime of validity of the underlying EFT.

\vskip 0.1 cm
{\bf Acknowledgements}: 
It is a pleasure to thank Riccardo Barbieri, Paolo Creminelli, Gia Dvali, Gregory Gabadadze, Oriol Pujolas and Filippo Vernizzi for valuable discussions. The work of D.P. is supported in part by MIUR-FIRB grant RBFR12H1MW and by funds provided by Scuola Normale Superiore through the program "Progetti di Ricerca per Giovani Ricercatori". E.T. is supported in part by MIUR-FIRB grant RBFR12H1MW. The work of P.U. is supported in part by the DOE grant DE- SC0011784.

\appendix

\section{The EFT for cosmological perturbations}
\label{appA}

In this appendix we summarize some of the technical details on computing the two point function of adiabatic perturbations on NEC-violating cosmological backgrounds. We closely follow the discussion of \cite{Creminelli:2006xe, Creminelli:2010ba}, generalizing the relevant expressions found in those references whenever appropriate.

\subsection*{Galileons in ADM variables}

It will prove convenient to work in the $(3+1)$ form of our generalized galileon action \eqref{ggg}. The necessary expressions have been derived in \cite{Creminelli:2010ba}, and we just summarize their results, with a minimal amount of adjustment relevant to our case. 
The $(3+1)$ decomposition of spacetime \cite{Arnowitt:1962hi} yields the following form for the four-dimensional metric
\beq
ds^2=-N^2 dt^2+g_{ij}(N^i dt+dx^i)(N^j dt+dx^j)~,
\eeq 
where $N\equiv 1/\sqrt{-g^{00}}$ and $N^i$ are the standard lapse and shift variables, while $g_{ij}$ is the induced metric on equal-time hypersurfaces (its determinant denoted by $g_3$ in what follows).
In the \textit{unitary gauge} defined by the absence of $\pi$-perturbations, $\pi(x,t)=\pi_0(t)$, the full action \eqref{ggg} can be written in terms of these variables in the following way (see \cite{Creminelli:2010ba} for derivation in the case of galilean genesis, $\mathcal{F}_1=e^{2\pi_0}$ and $\mathcal{F}_2=1$)
\ba
\label{adm}
S&=&S_g+S_\pi\\
S_g&=&\frac{1}{2}\mpl^2 \int d^4 x~\sqrt{g_3}N\big[R_3+(K^{ij}K_{ij}-K^{i~2}_{~i}) \big]\\
S_{\pi}&=&f^2 ~\int d^4 x ~\sqrt{g_3}N \bigg [-\mathcal{F}_1\(\pi_0\)\frac{\dot \pi_0^2}{N^2} + \frac{4 \dot \pi_0^3}{9 H_0^2} \frac{1}{N^3}K^i_{~i} + \mathcal{F}_2\(\pi_0\)\frac{\dot \pi_0^4}{3 H_0^2}\frac{1}{N^4}\bigg],
\ea
where $K_{ij}$ denotes the extrinsic curvature of equal-time hypersurfaces
\beq
K_{ij}=\frac{1}{2N} [\dot g_{ij}-\nabla_i N_j-\nabla_j N_i]~.
\eeq

\subsection*{Effective field theory}

To study the scalar spectrum of \eqref{adm}, one can readily employ the standard EFT of inflation formalism \cite{Creminelli:2006xe, Cheung:2007st}. The lagrangian for metric perturbations is largely constrained by the requirement of nonlinearly realizing spontaneously broken time translations, while 3D rotations as well as time- and space-dependent spatial diffs $x^i\to x^i+\xi^i(t,\vec{x})$ are realized linearly. The generic matter (non-Einstein-Hilbert) action can then be written as in \eqref{s_pi}
where we have defined $E_{ij}\equiv NK_{ij}$ and $\delta E^i_{~j}=E^i_{~j}- H \delta^i_{j}$. The first line gives the only terms that start linearly in metric perturbations, therefore their coefficients are completely fixed by the background equations. On the other hand, the second line contains terms that are manifestly at least quadratic in perturbations, their coefficients \textit{a priori} unconstrained. We have written out four such terms up to the quadratic order in derivatives. While only the first two are generated at the 'classical' level by the action \eqref{ggg},
\beq
\label{M's}
M^4(t)=\frac{4}{3} \frac{f^2}{H_0^2} \(2 \mathcal{F}_2\(\pi_0\) \dot\pi_0^4+\dot \pi_0^2\ddot\pi_0+9 H\dot\pi_0^3\), \quad \hat M_3^3(t)=\frac{4}{3}\frac{f^2}{H_0^2}\dot\pi_0^3~, 
\eeq
the rest of the operators are expected to be present ($\bar M',\bar M \neq 0$), although suppressed by whatever the quantum expansion parameter of the theory is. 

The unbroken spatial diffs can always be fixed so as to put the $3D$ metric in the follwing form \cite{Maldacena:2002vr}
\beq
g_{ij}=a^2(t)[(1+2\zeta)\delta_{ij}+\gamma_{ij}],\qquad \p_i\gamma_{ij}=\gamma_{ii}=0~,
\eeq
where $\zeta$ and $\gamma$ capture the physical scalar and tensor perturbations. Modes of different helicity do not mix on the homogeneous and isotropic backgrounds considered here, so that one can discard the helicity-1 part of the shift altogether, setting $N^i\equiv \delta^{ij}\p_j\beta$. 

\subsection*{Generalized galileons} 

In the effective theory for perturbations \eqref{s_pi}, both the lapse and the shift can be integrated out from their respective equations of motion\footnote{For the purposes of deriving the quadratic action for $\zeta$ only the linearized version of these equations matters. This is because  the contribution of higher-order terms in $\delta N$ and $\beta$ to the free  action for the curvature perturbation vanishes on-shell.}  ($\delta N\equiv N-1$):
\ba
\label{lapseeq}
\mpl^2 \bigg[R_3-\frac{1}{N^2}(E^{ij}E_{ij}-E^{i~2}_{~i})+\frac{2}{N^2}\dot H-2(3H^2+\dot H)\bigg]+2M^4\delta N-2\hat M^3\delta E^i_{~i}=0,~~~~~~\\
\label{shifteq}
\nabla_i\bigg[\mpl^2 \frac{1}{N} \(E^i_{~j}-\delta^i_j E^k_{~k}\) -\delta^i_j \hat M_3^3 N\delta N-\bar M'^2 N \delta E^i_{~j}-\delta^i_j\bar M^2 N\delta E^k_{~k}  \bigg]=0.~~~~~~
\ea

Let us start with the case of $\bar M'=\bar M=0$, relevant for the generalized galileon lagrangian \eqref{ggg}.
Upon expanding to the linear order in perturbations, the above equations can be algebraically solved for $\delta N$ and $\beta$ in terms of $\zeta$
\ba
\label{lapse}
\delta N&=&\frac{2\mpl^2}{2\mpl^2 H-\hat M^3}\dot \zeta \\
\nabla^2\beta &=&-\frac{2\mpl^2}{2\mpl^2 H-\hat M^3}~\frac{1}{a^2}\nabla^2\zeta +\frac{-4\mpl^4\dot H-12\mpl^2 H\hat M^3+3 \hat M^6+2\mpl^2 M^4}{(2\mpl^2 H-\hat M^3)^2}~\dot\zeta.~~~~~~~
\ea
Having obtained the lapse and shift, one can plug their expressions back into \eqref{s_pi}, that, after a few integrations by parts finally yields the quadratic action for $\zeta$
\beq
S_\zeta=\int d^4x~ a^3~\bigg[A(t)~\dot\zeta^2-B(t)~\frac{1}{a^2}\(\vec{\nabla}\zeta\)^2   \bigg]~,
\eeq
where the kinetic coefficients are given by the following expressions
\ba
A(t) &=&\frac{\mpl^2 (-4 \mpl^4 \dot H-12\mpl^2 H \hat M^3+3\hat M^6+2\mpl^2 M^4)}{(2\mpl^2H-\hat M^3)^2}\\
B(t)&=&\frac{\mpl^2 \(-4 \mpl^4 \dot H+2\mpl^2 H \hat M^3-\hat M^6+2\mpl^2\p_t\hat M^3\)}{(2\mpl^2H-\hat M^3)^2}~.
\ea
The speed of sound of the (short-wavelength) curvature perturbation $\zeta$ is 
\beq
c_s^2(t)=\frac{B(t)}{A(t)}~.
\eeq
These expressions have been used in Sec. \ref{analytic} and Appendix \ref{num} in testing for stability of solutions with the genesis - dS transition. 

\subsection*{Effects of higher (spatial) derivative operators} 
 
In order to assess the role of higher-order operators, we generalize the previous calculation to the case of non-zero $\bar M'$ and $\bar M$.
Linearizing and solving \eqref{lapseeq} and \eqref{shifteq}, one finds 
\beq
\delta N=\frac{2 (\mpl^2-\bar M'^2) (2 \mpl^2 H-\hat M^3 )\dot \zeta+ 2\mpl^2 (\bar M^2+\bar M'^2)\nabla^2\zeta/a^2}{2\mpl^2 H^2 (2 \mpl^2-3\bar M^2-3\bar M'^2)+(\bar M^2+\bar M'^2) (M^4-2\mpl^2\dot H)+\hat M^3 (\hat M^3-4\mpl^2 H)},~~\nn \\ \nn\\
\nabla^2\beta =\frac{(2\mpl^2+3\bar M^2+\bar M'^2) (M^4-2\mpl^2 \dot H)-6\mpl^2 H^2(3\bar M^2+\bar M'^2) +3 \hat M^3 (\hat M^3-4 \mpl^2 H)}{2\mpl^2 H^2 (2 \mpl^2-3\bar M^2-3\bar M'^2)+(\bar M^2+\bar M'^2) (M^4-2\mpl^2\dot H)+\hat M^3 (\hat M^3-4\mpl^2 H)}\dot\zeta ~~\nn \\
-\frac{2\mpl^2(2 \mpl^2 H-\hat M^3)}{2\mpl^2 H^2 (2 \mpl^2-3\bar M^2-3\bar M'^2)+(\bar M^2+\bar M'^2) (M^4-2\mpl^2\dot H)+\hat M^3 (\hat M^3-4\mpl^2 H)}\frac{\nabla^2 \zeta}{a^2}.~~ \nn
\eeq
Upon substitution back into \eqref{s_pi}, this yields the following quadratic action for the curvature perturbation 
\beq
S_\zeta=\int d^4x~ a^3~\bigg[A(t)~\dot\zeta^2-B(t)~\frac{1}{a^2}\(\vec{\nabla}\zeta\)^2-C(t)~\frac{1}{a^4}\(\vec{\nabla}^2\zeta\)^2 \bigg]~,
\eeq
with the kinetic coefficients 
\ba
\label{A'}
A &=&(\mpl^2-\bar M'^2)\cdot X\\
\label{B'}
B&=&-\mpl^2-\frac{1}{a}~\p_t Y ~, \\
\label{C'}
C&=&\frac{2 \mpl^4 (\bar M^2+\bar M'^2)}{Z} ~,
\ea
where the three auxiliary functions $X,~Y$ and $Z$ have been defined as follows
\ba
X&=&\frac{(2\mpl^2+3\bar M^2+\bar M'^2) (M^4-2\mpl^2 \dot H)-6\mpl^2 H^2 (3\bar M^2+\bar M'^2)+3 \hat M^3 (\hat M^3-4 \mpl^2 H)}{2\mpl^2 H^2 (2 \mpl^2-3\bar M^2-3\bar M'^2)+(\bar M^2+\bar M'^2) (M^4-2\mpl^2\dot H)+\hat M^3 (\hat M^3-4\mpl^2 H)}~,\nn \\ \nn\\
Y&=&a\cdot \frac{2 \mpl^2 (\mpl^2-\bar M'^2) (\hat M^3-2\mpl^2 H)}{2\mpl^2 H^2 (2 \mpl^2-3\bar M^2-3\bar M'^2)+(\bar M^2+\bar M'^2) (M^4-2\mpl^2\dot H)+\hat M^3 (\hat M^3-4\mpl^2 H)},\nn\\\nn\\
Z&=&2\mpl^2 H^2 (2 \mpl^2-3\bar M^2-3\bar M'^2)+(\bar M^2+\bar M'^2) (M^4-2\mpl^2\dot H)+\hat M^3 (\hat M^3-4\mpl^2 H)~.\nn
\ea

\section{Extended genesis: numerical study}
\label{num}

In this appendix we study an explicit illustrative model possessing cosmological solutions with the genesis-dS transition. In principle, any theory described by $S_{1,2}$ of Sec. \ref{sec2} for $e^\pi\ll1$ and $e^\pi\gg1$ is expected to do the job of reproducing the extended genesis cosmologies. Perhaps the simplest example is provided by \eqref{simplemodel}. For $\beta=0$ the theory is just the conformal galileon and when starting out in the GG phase, the expansion rate of the universe  diverges and the background exits the regime of validity of the EFT at some finite time, the scalar profile gradually growing as $e^\pi\sim 1/t$. 
For a nonzero $\beta$ however, the dynamics of the system is completely altered as soon as $\beta e^{2\pi}$ becomes of order, or greater than unity: the theory becomes effectively described by a $P(X)$ - type lagrangian with a cubic galileon self-interaction -- resulting, as we will show shortly, in  transition into an inflationary de Sitter phase. 

\begin{figure}[t]
\centering
\includegraphics[width=.38\textwidth]{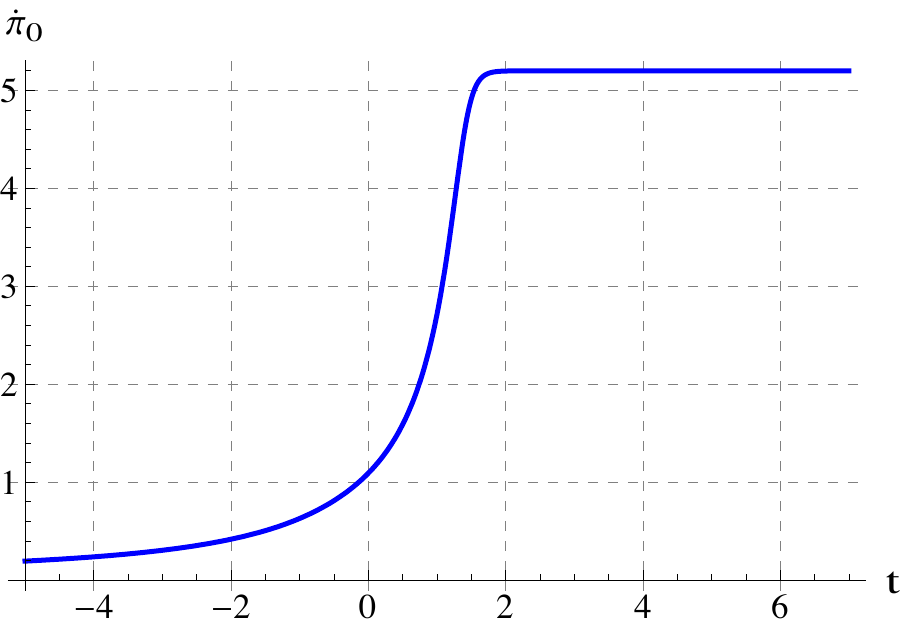} \quad
\includegraphics[width=.38\textwidth]{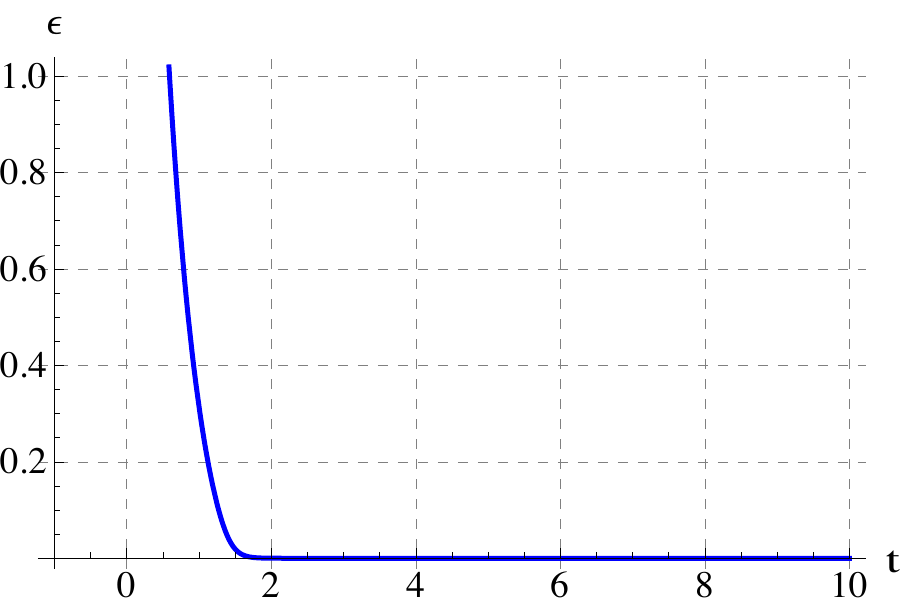}\\
\includegraphics[width=.38\textwidth]{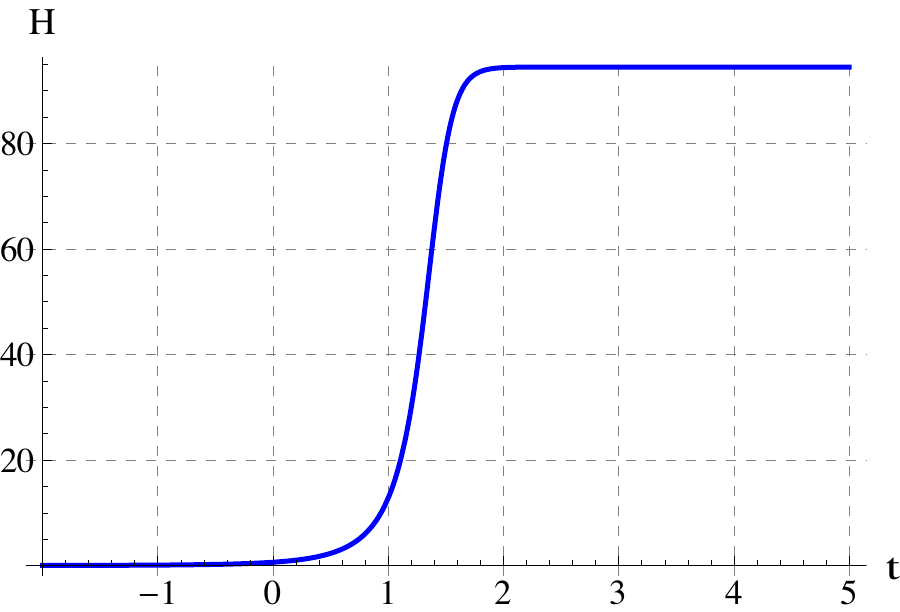}\quad
\includegraphics[width=.41\textwidth]{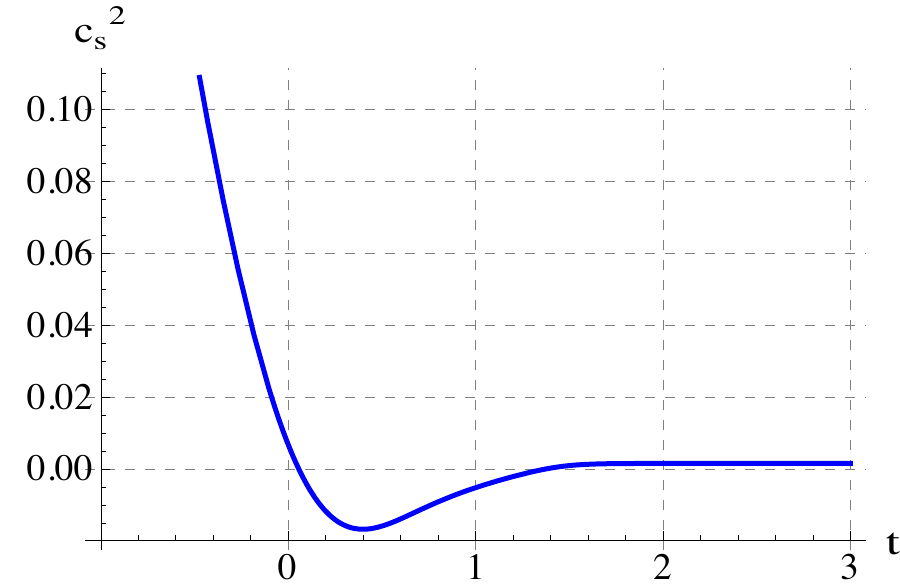}
\caption{Numerical solutions to the theory \eqref{simplemodel}, exhibiting the genesis - dS transition. Shown are the time evolution of quantities $\dot\pi_0,~ \varepsilon=\dot H/H^2,~H$ and the squared speed of sound $c_s^2$ of the curvature perturbation $\zeta$ (for early enough times not displayed in the plots, $c_s^2$ asymptotes to one as required by galilean genesis \cite{Creminelli:2010ba}). }
\label{figuretwo}
\end{figure}

Fig. \ref{figuretwo} illustrates a typical solution from our numerical study, obtained by integrating expressions for $\dot H$ and $\ddot \pi$ with the initial conditions, relevant for galilean genesis. We have assumed $\beta=0.001$ and $\mpl=f=10^6\cdot H_0$, setting $H_0$ (related to $\Lambda$ as in \eqref{ggsol}) as the unit mass scale.
Shown are the (time dependent) background quantities $\dot\pi_0,~ \varepsilon=\dot H/H^2,~H$ and the squared speed of sound $c_s^2$ of the curvature perturbation $\zeta$. For early enough times not displayed in the plots, $c_s^2$ asymptotes to 1 due to the emergent conformal symmetry. The graphs for the Hubble rate and the time derivative of $\pi_0$ clearly show the genesis - de Sitter transition, the scalar field acquiring a liear $\pi\propto t$ profile at late times. 

An explicit computation of the quadratic $\zeta$ action for the theory \eqref{simplemodel} is carried out in detail in appendix \ref{appA}.
While complete stability and (sub)luminality of the given backgrounds can be readily checked analytically at both asymptotics, the short transition region between 
the two phases displays gradient instability, at least for the values of parameters that we have been able to cover in numerical studies (we have checked explicitly that for all considered solutions, the flip of sign of the $c_s^2$ quantity stems from the gradient energy becoming negative - not the kinetic one, that would lead to a more severe ghost instability). 
For all solutions displaying the genesis - de Sitter transition, the squared speed of sound varies from unity at early times (as required by galilean genesis \cite{Creminelli:2010ba}) to a small value $c_s^2\lsim 0.03$ in the asymptotic future\footnote{It can be shown completely analytically \cite{Kobayashi:2010cm}, that for the most general galileon theory of the form \eqref{ginf}, the speed of sound for the scalar perturbation on a de Sitter background is bounded from above, $c_s^2\leq 0.031$ -- precisely what we are finding numerically for the extended genesis' future asymptotics.}, via a slight dip below zero in between that lasts from a few Hubble times to a fraction thereof -- depending on a solution. 
In principle, gradient instability has a characteristic time scale of order at least the quantum cutoff of the theory, therefore a background with this feature can not be considered fully legitimate.  

We present theories admitting fully stable cosmologies with the genesis-dS transition in Sec. \ref{analytic}, but note that even for the present simple theory the small (order per-cent) negative squared speed of sound corresponding to the gradient instability of Fig. \ref{figuretwo} suggests that it can be naturally cured by incorporating higher-order corrections in the effective theory for perturbations \cite{Creminelli:2006xe, Cheung:2007st}. While we carry out a systematic study of higher-order effects in Sec. \ref{hdim} (see also Appendix \ref{appA}), let us give a quick argument here. The interplay between higher (spatial) derivative operators contributing $\sim k^4$ terms to the IR dispersion relation for the scalar perturbation, and the presence of the cosmological horizon can stabilize the system against potential gradient instabilitiy (see, e.g., Ref.\cite{Creminelli:2006xe}).  This can be seen as follows.
At the level of four derivatives, one can add to the effective theory for perturbations on our background solution of Fig. \ref{figuretwo} the following term\footnote{There are other, more relevant terms beyond the leading order in the EFT, see Sec. \ref{hdim} for a systematic study. However, for the illustrative purposes we are after, we neglect them here.}  
\beq
\Delta S\sim \int d^4x~\sqrt{-g} ~\kappa(t)~ R_3^{~2}~,
\eeq 
where $R_3$ is the scalar curvature of the three-dimensonal metric, induced on equal time hypersurfaces and $\kappa$ is an arbitrary dimensionless time-dependent coupling. This term adds a higher-spatial derivative contribution to  the (unitary gauge) quadratic action for the curvature perturbation
\beq
\Delta S_\zeta \sim -\int d^4 x ~ \kappa~ \frac{1}{a}~(\vec{\nabla}^2\zeta)^2~.
\eeq 
Since $\kappa$ is an arbitrary function, it can always be chosen so as to render the instability scale for the background solution of Fig. \ref{figuretwo} smaller than the relevant instantaneous Hubble rate. Indeed, at frequencies larger than Hubble, the canonically-normalized curvature perturbation is described by the following action
\beq
S_\zeta=\int d^4x~ a^3~\bigg[\dot\zeta_c^2-c_s^2~\frac{1}{a^2}\(\vec{\nabla}\zeta_c\)^2   -\frac{\kappa}{A}~\frac{1}{a^4}~(\vec{\nabla}^2\zeta_c)^2~ \bigg],
\eeq
where $c_s^2$ is negative in the region with gradient instability. At large enough (physical) momenta, $  k^2 \gsim |c_s^2| A/\kappa$, the system is stabilized by higher-order effects. Requiring the corresponding frequency to be less than the instantaneous Hubble rate then yields the condition on $\kappa$ for a completely stable background solution
\beq
\kappa(t) \gsim \frac{c_s(t)^4 A(t)}{H(t)^2}~.
\eeq
Note the strong dependence ($\propto c_s^4$) on the scalar speed of sound of the lower bound on the coefficient $\kappa$. In particular, for small $c_s^2$, one can expect higher-order effects to easily cure the leading-order gradient instability.


\bibliographystyle{utphys}
\addcontentsline{toc}{section}{References}
\bibliography{extendedgenesis}

\end{document}